# Multifunctional epoxy nanocomposites reinforced by two-dimensional materials: A review

**Ming Dong[1], Han Zhang[2], Lazaros Tzounis[3,4], Giovanni Santagiuliana[2], Emiliano Bilotti[2] and Dimitrios G. Papageorgiou[2*]**

[1] *School of Physics and Astronomy, Queen Mary University of London, London E1 4NS, United Kingdom*

[2] *School of Engineering and Materials Science, Queen Mary University of London, London E1 4NS, United Kingdom*

[3] *Department of Materials Science & Engineering, University of Ioannina, GR-45110 Ioannina, Greece*

[4] *Mechanical Engineering Department, Hellenic Mediterranean University, Estavromenos, 71004 Heraklion, Greece*

*Corresponding author's email: d.papageorgiou@qmul.ac.uk*

## Abstract

Epoxy resins are thermosetting polymers with an extensive set of applications such as anticorrosive coatings, adhesives, matrices for fibre reinforced composites and elements of electronic systems for automotive, aerospace and construction industries. The use of epoxy resins in many high-performance applications is often restricted by their brittle and flammable nature, the relatively low fracture toughness and poor thermal and electrical properties. Various two-dimensional (2D) materials, such as graphene (Gr), hexagonal boron nitride (h-BN), transition metal dichalcogenides and MXenes, provide vast opportunities to endow multifunctional properties and reinforce epoxy resins for advanced applications. In this review, the current literature status of epoxy nanocomposites reinforced with 2D materials has been thoroughly examined. The structures and intrinsic properties of epoxy resins and two-dimensional materials have been briefly summarized. Recent advances in the strategies of incorporating 2D materials into epoxy matrices have also been presented. Most importantly, the mechanical, tribological, thermal, electrical, flame retardant and anticorrosive properties of epoxy nanocomposites reinforced with 2D materials have been reviewed in detail. Finally, the current status of the field along with future perspectives have been discussed with regards to the effectiveness of various 2D nanofillers towards reinforcement.

## 1. Introduction

During the past decades, polymer nanocomposites have attracted huge attention from both academia and industry, due to their excellent specific mechanical properties, thermal stability, electrical/thermal conductivity and chemical resistance properties, amongst others [1, 2]. Thermosets and thermoplastics are two of the most important categories in the large family of polymeric materials that are commonly used as matrices in polymer composites. Compared to thermoplastics, thermosets are more favoured in advanced engineering applications, due to their high modulus and strength, alongside with their low viscosity and ease of manufacturing at room temperature. Amongst the list of thermosetting materials, epoxy resins possess superior characteristics and therefore consist probably the most widely used resins nowadays [3]. For example, epoxy resins exhibit high mechanical strength and hardness, due to their highly crosslinked rigid segments. Epoxy resins also develop low residual stress, due to relatively low levels of shrinkage during curing, while possessing good resistance to heat and chemicals [4]. Given that the demand for high performing lightweight structural materials has been continuously increasing, epoxy resins have been extensively used over the last decades especially in advanced fibre reinforced polymer (FRP) composites towards structural applications, i.e. in aerospace and automotive sectors, etc. [4].

Diverse microscale or nanoscale fillers have been incorporated into epoxy resins to further improve their mechanical properties and endow multifunctionality. Conventional fillers with microscale dimensions such as short glass/carbon fibres, carbon blacks, silica and metal micro particles have been widely used to reinforce epoxy resins. However, these traditional fillers display some unavoidable disadvantages [5]. For instance, high filler loadings are often required to achieve modest mechanical reinforcement, resulting in added weight penalty of the components. In addition, interfacial defects and weak interfacial interactions can result in catastrophic failure upon loading. To overcome these disadvantages, nanoscale fillers such as clay, silica ($SiO_2$), carbon nanotubes (CNTs) and graphene have been widely used as reinforcing agents in epoxy nanocomposites having a strong effect mainly in improving both static, dynamic and specifically the fracture toughness properties [6, 7]. These nanofillers can be divided into three categories: zero-dimensional (0D) nanoparticles or nanospheres, one-dimensional (1D) nanotubes or nanowires, and two-dimensional (2D) nanosheets. One-dimensional nanoparticles such as carbon nanotubes (CNTs) have attracted lots of attention, due to their excellent thermal, electrical, mechanical properties, and unique structures with very high aspect ratios. Single-walled carbon nanotubes (SWCNTs) are essentially rolled-up monolayer graphene sheets that display a Young's modulus of 1 TPa and tensile strength in the order of 100 GPa, thermal conductivity of ~6000 W $m^{-1}K^{-1}$ and high electrical conductivity in the range of $10^4$–$10^8$ S/m [8]. These characteristics render CNTs one of the most promising reinforcing fillers in epoxy nanocomposites, explaining the extensive research campaign conducted on CNT-reinforced polymer nanocomposites [9, 10].

It should be noted that the use of CNTs in nanocomposites also has some disadvantages compared to 2D nanofillers. For example, as it will be shown in detail later, 2D nanosheets with high surface to volume ratio are more favourable for reinforcement in polymers as random orientation of 2D materials reduces the composite modulus to almost half, while for 1D materials the composite modulus is reduced by a factor of 5 [11]. In addition, the 1D geometry of the CNTs makes them more prone to aggregation, while the entanglements between CNTs are very difficult to break resulting in a non-homogeneous dispersion within the epoxy matrix. Moreover, the 1D geometry of CNTs often results in extremely high viscosity due to entanglement (even at low wt.%, <1.0 wt.%), and this leads further in significant difficulties during processing i.e. mould filling and fibre infiltration during the manufacturing of composite materials via resin transfer moulding (RTM), resin infusion, or in a lower scale, during prepreg manufacturing. In contrast, polymer nanocomposites reinforced with 2D fillers are easier to process and achieve a better dispersion as a result of easier shear between the 2D nanosheets arising from weak interlayer forces. Additionally, the shape and the dimensions of 2D nanofillers are beneficial for constructing ordered structures that enable better stress transfer [12] and allow better barrier properties while maintaining multifunctionality. The entangled nanotubes on the other hand can act as stress concentration points during the application of strain and can even reduce the mechanical properties of a polymer matrix [13]. Consequently, 2D nanofillers, with their high aspect ratio, multifunctional properties and ease of processing hold huge potential in the reinforcement of epoxies and other polymeric matrices.

Amongst the 2D nanofillers, nanoclays (e.g. montmorillonite (MMT)) and graphene-related materials are two of the most widely used reinforcements for epoxy resins [6, 14]. Ever since its isolation in 2004 [15], graphene has been regarded as one of the most promising reinforcing agents in polymer nanocomposites, due to its exceptional properties that include high modulus and strength along with excellent thermal and electrical properties [16]. Meanwhile, other 2D materials beyond graphene such as hexagonal boron nitride (hBN), molybdenum disulphide ($MoS_2$) and MXenes, such as $Ti_3C_2T_x$, also display excellent mechanical properties and a unique combination of characteristics. For example, hBN shows very high thermal conductivity, similar to that of graphene, while contrary to graphene it is an electrical insulator, making it suitable for heat sink applications. Additionally, $MoS_2$ [17] is a chemically versatile 2D material, compared to the chemically-inert graphene, showing the largest piezoelectric activity among 2D materials [18]. In terms of mechanical properties, the elastic moduli of the three above-mentioned 2D materials are located between the values of nanoclay and graphene, at about 300-900 GPa [19]. From the above, it can be easily realised that a range of 2D nanofillers can be utilised for the efficient reinforcement of epoxies based on the targeted applications and properties. It is highly worthwhile then to gain an insight into fabrication and multifunctionality of epoxy nanocomposites reinforced with various 2D nanomaterials, as well as elucidate the structure-property relationships towards maximisation of reinforcement in future technologies.

This review summarizes recent advances in the preparation and the mechanical, tribological, thermal, electrical, flame retardant and anticorrosive properties of epoxy nanocomposites reinforced with 2D nanomaterials. Herein, we have focused on graphene-derived (e.g. graphene and graphene oxide (GO)) materials and on a number of other 2D materials (e.g. hBN, $MoS_2$, $Ti_3C_2T_x$) for the reinforcement of epoxy resins. The unique structure and properties of epoxies and some of the most important 2D materials are summarized in Section 2. In the following section (Section 3), different preparation methods to incorporate 2D materials into epoxy resins are presented. Finally, the role of different 2D materials in the reinforcement of various properties of epoxy nanocomposites are discussed with a focus on the attained "structure-property" relationship (Section 4).

## 2. Structure and properties of epoxy and two-dimensional materials

### 2.1 Epoxy

Epoxy resins were first synthesized in 1909 by Prileschajew [3] and they can be considered as low molecular weight pre-polymers containing more than one epoxide group. The epoxide groups can react with curing agents (hardeners) over a wide temperature range to crosslink and form cured epoxy resins. The final crosslinking density and the properties of epoxy resins vary with the types of resin i.e. chemistry, molecular architecture of the epoxide monomer, as well as the employed curing agents and curing conditions.

Curing agents play a vital role in the final properties of epoxy resins. Epoxy resins react with a curing agent to form a crosslinked network through the polymerisation/curing process. Different curing agents such as amine-type, alkali, anhydrides and catalytic curing agents have been used [4]. The curing process can be conducted at room temperature or at elevated temperatures. Normally, epoxy resins obtained by high-temperature curing exhibit better properties such as higher glass transition temperature, as well as higher stiffness and strength [6].

Epoxy resins have numerous salient characteristics and have been widely applied in different fields. Apart from the low residual stress, excellent mechanical properties and good resistance to heat and chemicals, various epoxy resin formulations also exhibit strong adhesion with substrates and have been used as adhesives for metals, aircrafts, automobiles [20]. In addition, epoxies can be used as coatings with anticorrosion properties. Despite the obvious advantages, some unavoidable disadvantages such as their brittle nature and their poor thermal and electrical properties have restricted the application of epoxy resins in several cases. In this aspect, various 2D materials known for their excellent mechanical, thermal, electrical and other multifunctional properties can be incorporated into epoxy resins to further promote their engineering applications.

## 2.2 Two dimensional (2D) materials

The research field of 2D materials keeps growing since the isolation of graphene in 2004; numerous 2D materials have been studied over the last decade. The structures of some of the most widely analysed 2D materials, including graphene, hBN, $MoS_2$, black phosphorus (BP), MXenes and metal organic framework (MOF) are presented in **Figure 1**. Most of the 2D materials possess high elastic modulus and fracture strength as a result of their strong in-plane covalent bonds, which means that they could potentially be effective mechanical reinforcing agents in epoxy nanocomposites. In this section, we will briefly summarize the structure and properties of 2D materials. Additionally, a comparative Table (Table 1) with the properties of the 2D materials that have been discussed herein is presented at the end of this Section.

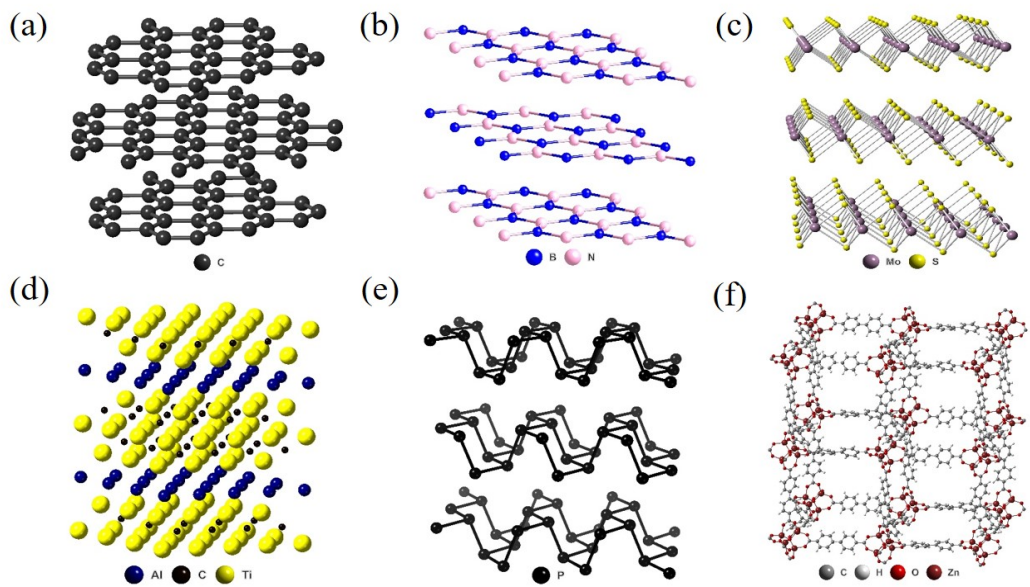

**Figure 1**. Structures of diverse 2D materials, including (a) graphene, (b) boron nitride, (c) $MoS_2$, (d) $Ti_3AlC_2$, (e) black phosphorus and (f) MOF-10.

*2.2.1 Graphene*

Graphene is composed of a single layer of carbon atoms arranged in a two-dimensional honeycomb structure as shown in **Figure 1**. Originating from its sp$^2$ hybridized bonds and π state bands, graphene exhibits unprecedented mechanical [21], thermal [22] and electrical [23] properties as summarized in **Table 1**. In addition, graphene also exhibits high intrinsic flexibility, large surface area and impermeability to gases and moisture [24]. Microscale structural superlubricity has also been achieved in bilayer graphene [25]. It should be noted that monolayer and two/three-layer graphene are not commonly used in polymer nanocomposites due to various difficulties associated with the bulk preparation of such nanomaterials and nanocomposites. The increase of graphene thickness results in significantly lower mechanical, thermal and electrical properties compared to the monolayer material, which are still adequate compared to other types of nanofillers. The geometrical characteristics, large aspect ratio and good processability are some of the unique attributes that make the family of graphene-related materials highly effective in reinforcing polymer nanocomposites.

Graphene derivatives are widely used as nanofillers in epoxy nanocomposites. Graphene oxide (GO) made by oxidation and exfoliation of graphite is one of the mostly used graphene derivatives in polymer nanocomposites. Compared with graphene, monolayer GO exhibits a lower elastic modulus due to the presence of lattice defects that originate from the oxidation process [26]. The defects also have a negative effect on the electrical and thermal conductivity of GO, thus hindering the application of GO in electrically and thermally conductive composites. However, the presence of oxygen functional groups such as epoxides, carboxylic acids and alcohols reveals that GO can interact chemically with certain polymers and especially with epoxy resins. This fact is certainly beneficial for improving the interfacial shear strength and stress transfer efficiency in GO-reinforced composites. It should be noted that the strong interactions between GO nanoplatelets and the polymer matrix can lead to a significant increase in viscosity. This in turn leads to processing difficulties, an inhomogeneous dispersion of the GO nanoflakes in the nanocomposites and subsequently a knock-down effect on the final properties. As a result, for the case of GO, a balance should be kept between the degree of oxidation, the optimal loading, the processability of the nanocomposite and its ultimate properties.

GO can be reduced (reduced graphene oxide, rGO) to graphene nanosheets. Compared with other preparation methods of graphene, the reduction of GO is low-cost, due to the use of graphite as raw material and suitable for large-scale production. The final properties of rGO depend on the degree of reduction. On the one hand, the removal of functional groups and defects can increase the intrinsic properties of GO. On the other hand, the removal of functional groups can decrease the interactions between the filler and the matrix. It has been shown that the removal of oxygen functional groups can induce nanosheet re-agglomeration and as a result, this process can reduce the aspect ratio, increase the thickness of the nanoplatelets and therefore reduce the mechanical reinforcing efficiency [27].

### 2.2.2 Boron nitride

Boron nitride, also called "white graphene", is structurally similar to graphene, with boron and nitride atoms arranged in a honeycomb lattice (**Figure 1**). The Young's modulus and tensile strength of monolayer hBN are shown in **Table 1** [28]. Quite importantly, the increase of layer number (up to nine layers) does not degrade its mechanical properties due to strong interlayer interactions and suppressed sliding tendency under strain – that is in contrast with what happens for graphene [28]. When subjected to strain, the $2p_z$ orbitals of BN localize the electronic density and the interlayer sliding energy will increase; on the contrary, the orbitals in graphene tend to overlap and the sliding energy will decrease to zero or take negative values. The relative independence of the mechanical properties on the layer number is an advantage of BN in terms of mechanical reinforcement in polymer composites, compared to graphene. Moreover, hBN also possesses excellent thermal properties [29]. Hexagonal BN is resistant to oxidation. It starts to oxidize only at 700 °C and can sustain temperatures up to 850 °C [30], which are much higher compared to monolayer graphene that reacts with oxygen already from 300 °C, and fully decomposes at 450 °C [31]. Another unique feature of BN is that it is electrically insulating. Therefore, BN can be used to create mechanically robust, thermally conductive and electrically insulating polymer nanocomposites. In addition, its electrical insulating nature makes BN more suitable as long-term barrier material compared to graphene, since conductive graphene may introduce galvanic corrosion phenomena [32].

### 2.2.3 Transition Metal Dichalcogenides (TMDCs)

Different from the structure of graphene and BN, in transition metal dichalcogenides (TMDCs) the transition metal atoms are sandwiched between two layers of chalcogen atoms as shown in **Figure 1**. Among the list of TMDCs, $MoS_2$ has been used to reinforce a number of polymers [33]. Crystalline $MoS_2$ is met in three structural polytypes; trigonal, hexagonal and rhombohedral. The most common polytype is the hexagonal one, and the bulk mineral molybdenite is used for the exfoliation of few-layer $MoS_2$. The Young's modulus and tensile strength of monolayer $MoS_2$ were measured to be 270 ± 100 GPa and 22 ± 4 GPa, respectively [34], indicating that $MoS_2$ can also be a quite effective mechanical reinforcing agent in polymer composites, however with lower reinforcing capabilities compared to BN and graphene due to the important differences in their intrinsic mechanical properties. In addition, superlubricity has also been observed in layered $MoS_2$ [35], which implies that $MoS_2$ can be used to enhance the tribological performance of polymer composites. In terms of its electrical properties, hexagonal $MoS_2$ is a semiconductor while trigonal $MoS_2$ is metallic [33]. Finally, the use of $MoS_2$ can also improve the flame retardant properties of polymer composites as a result of its barrier effect that retards the heat release rate and the catalytic effect that improves char yield and suppresses the production of smoke.

*2.2.4 MXenes*

MXenes ($M_{n+1}X_nT_x$, where M stands for a transition metal, X stands for C or N and T represents the surface functional groups) are transition metal carbides, nitrides and carbonitrides, with more than seventy types of MAX ($M_{n+1}AX_n$, where M is a transition metal, A is an A group element, and X is C and/or N and n = 1 to 3) phases already reported [36]. The family of MXenes displays a crystal structure similar to the one of MAX phase ceramics. The $Ti_3C_2T_x$ layers are connected by strong hydrogen bonds, so interlayer sliding is depressed and the increase of layer number does not affect its mechanical properties pronouncedly [37]. The similar effect has been observed in GO films due to hydrogen bonding and multilayer BN due to interlayer B-N interaction. In addition, MXenes show excellent electrical conductivity, electromagnetic interference (EMI) shielding performance and charge storage capability [38, 39]. For example, the EMI shielding effectiveness of a 45 μm $Ti_3C_2T_x$ film was measured to be 92 dB as a result of its high electrical conductivity (4600 S cm$^{-1}$) and the induced multiple internal reflections [38]. The surface functional (oxygen and/or fluorine) groups also provide MXenes with good hydrophilicity, ion conductivity and good compatibility with polymer molecules. These properties make MXenes excellent candidates as multifunctional reinforcements in polymer nanocomposites for a number of applications.

*2.2.5 Other 2D nanofillers*

Apart from the above mentioned 2D nanofillers, other 2D materials such as black phosphorus (BP), metal-organic frameworks (MOFs) and covalent organic frameworks (COFs) are also used as fillers in polymer nanocomposites. Different from most 2D materials, BP exhibits anisotropic mechanical, thermal and electrical properties, due to its puckering structure (**Figure 1**). Additionally, BP can be used as a lubricant additive since superlubricity has been observed in BP [40]. COFs (e.g. COF-5, COF-42 and COF-300) are crystalline porous materials covalently linked by organic building units where each layer is connected by van der Waals (vdW) forces [41]. MOFs (e.g. UiO-67, ZIF-8 and MOF-10) are crystalline porous compounds, where the metal ions or clusters are connected by coordinating organic ligands to form bulk crystals [42]. COFs and MOFs are effective towards improving the anticorrosive and flame retardant properties of epoxy nanocomposites, which will be discussed within the next sections.

**Table 1.** Physical properties of different 2D materials (for BP the / symbol indicates the properties along the armchair/zigzag directions).

| Properties | Graphene | BN | MoS$_2$ | Ti$_3$C$_2$T$_x$ | BP |
|---|---|---|---|---|---|
| Density (g/cm$^3$) | 1.9-2.3 | 2.1 | 5.06 | 4.21 | 2.34 |
| Young's modulus (GPa) | 1000 [21] | 865 [28] | 270 [34] | 330 [37] | 27/59 [43] |
| Tensile strength (GPa) | 130 [21] | 70.5 [28] | 22 [34] | 17.3 [37] | 2.3/4.8 [43] |
| Thermal conductivity (Wm$^{-1}$K$^{-1}$) | 5000 [22] | 751 [29] | 34.5 [44] | – | 34/86 [45] |
| Thermal oxidation in air (°C) | 300-450 [31] | 700-850 [30] | 310 [46] | – | – |
| Electrical conductivity (S/m) | $6 \times 10^5$ [23] | – | – | $1.1 \times 10^4$ [39] | – |

# 3 Preparation of epoxy nanocomposites with 2D nanomaterials

Apart from the intrinsic properties of 2D nanofillers, the processing methods also play a vital role towards the final performance of epoxy nanocomposites. Solution blending via conventional mechanical stirring and/or sonication approaches, as well as high shear mechanical mixing are among the most frequently adopted methods to prepare epoxy nanocomposites considering their simplicity and versatility. Other methods to further promote the homogeneous distribution and/or tailored network of 2D nanofillers such as epoxy impregnation have also been suggested to fabricate epoxy nanocomposites. A comparison between different preparation methods is presented in **Table 2**.

## 3.1 Solution blending

Solution blending is a simple, versatile and effective method to prepare epoxy nanocomposites and has been used to incorporate various 2D nanofillers into epoxy resins. A typical solution blending process for the production of epoxy nanocomposites involves the dispersion of the 2D filler within an appropriate solvent and then the liquid resin is added to the suspension. Once the solvent is evaporated, the curing agent is added to the suspension and the mixture is commonly casted in a mould and cured. Despite its agility, the method also has some drawbacks. At large filler contents (> 10-20 wt%), achieving a homogeneous dispersion can be quite challenging. Additionally, the complete removal of the solvent is essential as any residue can degrade the properties of the nanocomposite while the use of large quantities of solvents can cause environmental concerns associated with their disposal. Finally, re-aggregation or restacking of 2D nanoplatelets is quite common during the blending step of the epoxies with the fillers.

*3.2 Mechanical mixing*

Mechanical mixing is another widely used processing method for the fabrication of epoxy nanocomposites, utilising high shear forces to disperse and/or exfoliate 2D fillers in the resin. Apart from high shear mixing, a typical example is the three-roll milling (TRM) process, where high shear forces between rollers can disperse 2D fillers such as graphene or hBN within the epoxy resins [47]. Li *et al*. [47] successfully fabricated epoxy/graphene nanocomposites by exfoliation of natural graphite to graphene nanoplatelets *in situ* during the TRM process, without using any solvents or additives. Graphene nanoplatelets (GNPs) with thickness in the range of 5-17 nm and with an aspect ratio of 300-1000 were obtained from natural graphite directly, and the epoxy nanocomposites without any separation/purification steps displayed enhanced mechanical and electrical properties [47, 48]. Other properties such as the fracture toughness ($K_{IC}$) of nanocomposites can be also enhanced as a result of tilting and twisting of the cracks due to the presence of the 2D nanofillers [49]. Chandrasekaran *et al.* [49] compared the effect of GNP dispersion within an epoxy resin by TRM and a process of combined sonication and high speed shear mixing. The authors observed an increased $K_{IC}$ and a more homogeneous distribution of GNPs by the TRM method, as a result of the high shear forces that are developed during TRM.

A common problem of mechanical mixing is that (similarly to solution blending) high filler contents (e.g. > 10-20 wt%) are not easily well-dispersed, resulting in aggregation phenomena. To improve the dispersion and alignment of 2D nanofillers in an epoxy matrix, a method that utilises compression moulding has been proposed in the literature, (schematically depicted in **Figure 2)** [50-52]. The method allows the creation of oriented filler networks and the closure of gaps between adjacent fillers, creating highly conductive pathways. This is another typical example of the importance of filler orientation towards the maximization of the ultimate properties of nanocomposites across a certain direction.

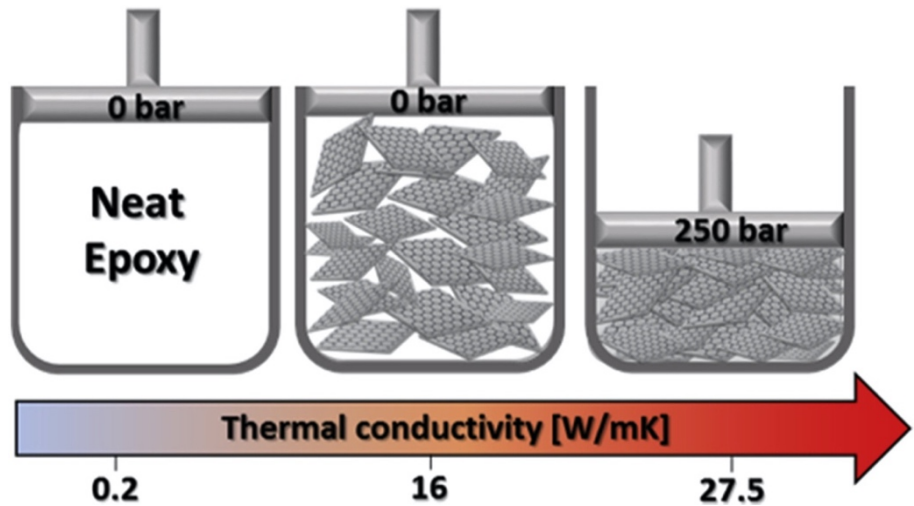

**Figure 2**. Compression method for the preparation of epoxy/GNP nanocomposites. Reproduced with permission from [52]. Copyright 2018 Elsevier.

In general, the reduction in the lateral dimensions of 2D fillers during mixing should be avoided or minimised, to preserve their high aspect ratio that is crucial for the resulting mechanical, electrical, thermal and functional properties endowed to the final nanocomposites. When the filler loading is high, a drastic increase in the resin viscosity is usually observed. In this case, the mechanical mixing methods might introduce excessive shear forces that reduce the filler lateral dimensions. The viscosity increase is also commonly accelerated by 2D fillers which already have functional groups on their surface (such as GO and MXenes) or by chemically functionalised fillers. It is worth noting that the subsequent processing steps are also highly important in determining final properties, as the dispersed nanofillers tend to re-agglomerate, due to their large surface area until the curing process is completed [53].

*3.3 Epoxy impregnation*

Epoxy nanocomposites can be also prepared by the epoxy impregnation process. Before the epoxy impregnation, 2D nanosheets are assembled into aligned 2D or 3D structures by vacuum filtration [54], freeze drying [55] or by other directed assembly methods [56]. Then, the epoxy resin is impregnated into the assembled structures to obtain epoxy nanocomposites. Compared to solution or mechanical blending methods, a well-ordered, aligned structure or a 3D network of nanofillers can be achieved by this method. Im *et al*. [54] fabricated hybrid epoxy nanocomposites reinforced with GO and multi-walled carbon nanotubes (MWCNT) using this method as shown in **Figure 3**. The GO and MWCNT mixture was filtered via vacuum filtration to prepare the composite cake and the epoxy resin along with the curing agent were deposited onto the cake to prepare the nanocomposites. The epoxy impregnation method also has some limitations. For example, freeze casting and vacuum filtration processes are time-consuming and quite challenging to be scaled up [56]. There is also difficulty in controlling the 2D nanofillers to follow the ice morphology precisely during the freeze drying process.

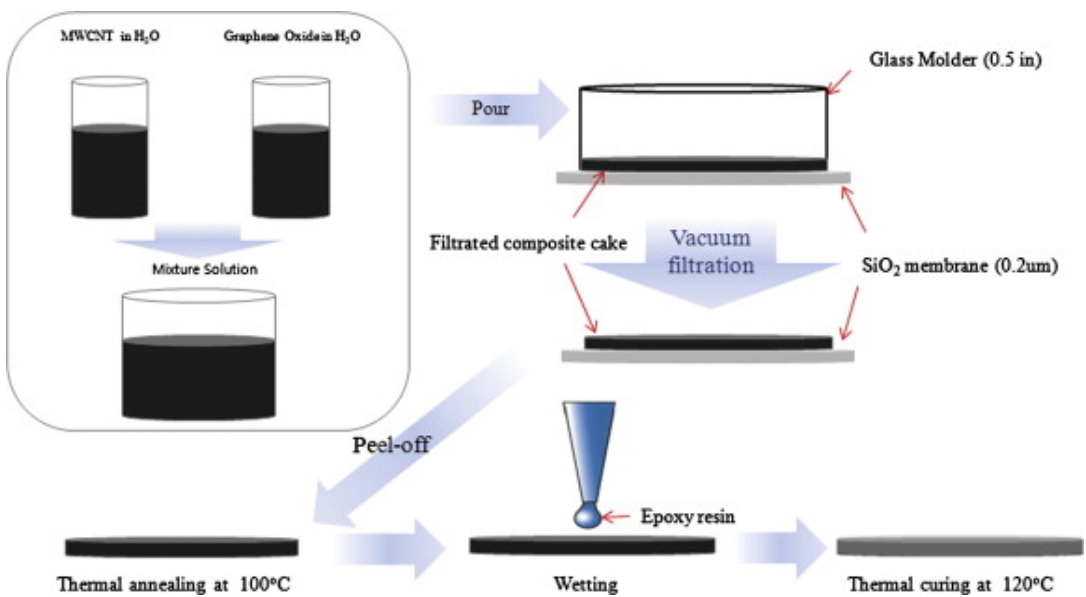

**Figure 3**. Vacuum filtration and epoxy impregnation for the preparation of GO/MWCNT/epoxy nanocomposites. Reproduced with permission from [54]. Copyright 2012 Elsevier.

**Table 2.** Comparison of different methods for the preparation of epoxy nanocomposites.

| Methods | | Advantages | Disadvantages |
|---|---|---|---|
| **Solution blending** | Casting | • Simple, effective and versatile | • Hard to control filler orientation |
| | | | • Poor dispersion at high filler content |
| **Mechanical mixing** | Casting | • Solvent-free | • Reduction in filler lateral size |
| | | • Easy operation | • Poor dispersion at high filler contents |
| | Hot pressing | • Controlled filler orientation | • High filler contents required to observe significant differences |
| **Epoxy impregnation** | Vacuum | • Controlled filler orientation | • Difficult to scale up |
| | | • Easy operation | |
| | Freeze casting | • Controlled filler orientation | • Energy- and time-consuming |
| | | • Allows high filler fractions | |
| | Hydrogel casting | • Cost-effective, large-scale size | • Difficult to control the layered structure |

## 4 Properties of epoxy nanocomposites reinforced with 2D materials

### 4.1 Mechanical properties

Compared to traditional materials such as steel, aluminium and titanium, a huge advantage of epoxy-based matrix composites is the combination of high strength and low weight. Therefore, they have been widely applied in the automotive industry as frame parts, chassis parts, floor panels, firewalls, seat structure, closure of trucks, and so on. Additionally, the high strength of pure epoxy resins can be utilised in structural materials for aeronautics and aerospace applications subjected to harsh environments such as extreme heat or cold and radiation. Epoxy-based composites have been used for the skins of empennages, wings and fuselages. Various 2D materials have been utilised to reinforce epoxy resins considering their excellent mechanical properties and high surface to volume ratio. These 2D materials can trigger toughening mechanisms and improve the elastic modulus, strength and toughness of epoxy resins significantly. It has been shown that the classical composite micromechanics are able to describe the mechanical reinforcement in polymer composites from 2D nanofillers [2, 57]. In the current section, the micromechanical modelling of epoxies reinforced by 2D materials will not be discussed in great detail, as this task goes beyond the scope of this review, but we will briefly summarise a few theories that have been recently proposed for the evaluation of reinforcement characteristics of the nanocomposites.

The Young's modulus of bulk nanocomposites, $E_c$, can be described by the modified rule of mixtures as

$$E_c = \eta_o \eta_l E_{eff} V_f + E_m V_m \qquad (1)$$

where $V_f$ and $V_m$ ($V_f + V_m = 1$) are the volume fractions of the filler and the matrix, $E_m$ is the modulus of the matrix and $E_{eff}$ is the effective modulus of the filler, $\eta_o$ is the Krenchel orientation factor ($\eta_o = 1$ for aligned nanosheets and 8/15 for randomly oriented ones [58]), and $\eta_l$ is the length efficiency factor that depends on the aspect ratio of the nanoplatelets. The spatial orientation is of utmost importance for the maximisation of mechanical reinforcement when the fillers are oriented along the direction of strain, as the amount of interface that contributes to the mechanical properties is maximised and better stress transfer is achieved. Quite significantly, the Krenchel factor for 2D materials is 8/15, which means that a random orientation of 2D fillers reduces the stiffness of the nanocomposites by less than a factor of 2. On the contrary, 3D (randomly) oriented fibres and nanotubes display a Krenchel factor 1/5, revealing that the stiffness, in this case, is reduced by a factor of 5. This shows that from a geometrical consideration point of view, a better degree of reinforcement should be achieved from randomly oriented 2D materials, compared to their randomly oriented 1D counterparts.

However, the final mechanical performance of polymer composites does not depend only on the aspect ratio, orientation, volume fraction, and effective modulus of the fillers, but also on other factors that are not represented in the modified rule of mixtures. Indeed, equation (1) is valid for composites containing perfectly dispersed fillers, but this is not always the case, especially for nanocomposites where nanofillers can agglomerate at low volume fractions (<2 vol%), due to their high aspect ratio and inter-particle attraction. Agglomerated nanoparticles are not in full contact with the polymer matrix, thus the external load cannot be efficiently transferred from the matrix as expected by Cox's shear-lag theory [59], and the nominal volume fraction of the nanofiller can only partially contribute to the mechanical reinforcement. A few micromechanical theories have been proposed in the literature in an attempt to account for agglomeration phenomena in polymer nanocomposites. Santagiuliana *et al.* [60] proposed that the reduced volume fraction (or effective volume fraction, $V_{f(eff)}$) effectively reinforcing a nanocomposite, that unavoidably contains agglomerates, can be calculated using a nanofiller dispersion level, $D$, from the following simple equation; $V_{f(eff)} = D \cdot V_f$ (with $0 \leq D \leq 1$), and experimentally found that $D$ is proportional to the ratio between the nanofiller-matrix contact area and the total nanofiller surface area. Since there is a maximum limit to the nanofiller-matrix contact area that is given by the maximum specific surface area that a polymer can reach [61], $D$ will always decrease after a critical nanofiller concentration (the onset of agglomeration), which explains the decreasing reinforcing efficiency commonly observed in nanocomposites with the increased amount of nanofiller. Similarly, Li *et al.* [62] proposed the introduction of an agglomeration factor ($\eta_a$) within the modified rule of mixtures (Eq. 1) that takes values between 0 and 1, where for $\eta_a = 0$ all flakes are aggregated and nearly no stress transfer takes place, while for $\eta_a = 0$ all flakes are well dispersed and interact strongly with the matrix to allow good stress transfer.

The shear modulus of the polymeric matrix and the interfacial shear strength also play a vital role in the mechanical performance of polymer composites according to the shear-lag theory [57]. This has been clearly demonstrated in the paper of Young *et al.* [57], where the authors studied a number of polymeric matrices of various stiffnesses reinforced with graphene nanoplatelets to understand the mechanics of reinforcement of polymer nanocomposites. According to the shear-lag theory for the deformation of individual graphene nanoplatelets, the filler modulus can be obtained from the following relationship:

$$E_f \approx \eta_o \frac{s^2}{6} \frac{t}{T} G_m \qquad (2)$$

where $t/T$ is the interfacial parameter, related to the proximity of neighbouring nanoparticles. Hence, for the simple case of nanoplatelets stacked between a polymer layer it can be considered that $t/T \sim V_f$. Then, by substituting into the rule of mixtures, the composite modulus can be given by the simple equation:

$$E_c \approx E_m \left[ 1 - V_f + \frac{s^2}{12} \frac{\eta_o}{(1+\nu)} V_f^2 \right] \qquad (3)$$

where the strong dependence of the $E_c$ on the square of the aspect ratio ($s^2$) and the volume fraction ($V_f^2$), along with the degree of orientation ($\eta_o$) and the interfacial parameter ($t/T$) can be seen. This theory gives a convincing explanation why the exceptional properties of 2D materials are difficult to be realised in soft matrices as a result of the ineffective stress transfer due to the low shear matrix modulus. Quite importantly, it shows that for most polymers the aspect ratio of the nanoplatelets and their orientation can be more important than their intrinsic modulus, therefore making other 2D materials besides graphene equally attractive for mechanical reinforcement [57].

Besides micromechanics, first principle calculations and molecular dynamics can be employed to evaluate the reinforcing mechanisms of polymers by 2D materials. Odegard's group has recently reported on the molecular-level structure of the interface between resins or hardeners, reinforced with graphene and boron nitride monolayers, by density functional theory [63]. The first principles calculations revealed that the interfacial adhesion was dependent upon the orientation of the resin on graphene, as a result of the van der Waals interactions. Additionally, a small degree of polarity at the interface can help in improving the mechanical properties of the nanocomposite. The same group performed molecular dynamic simulations on epoxy reinforced with GNPs, GO and functionalised GO (fGO) [64]. The results showed that the wrinkles on the GO materials improve the filler-matrix interlocking mechanism, leading to an increase in the out-of-plane shar modulus, while as expected, the mechanical properties of GO and fGO are lower than that of the GNPs as a result of the transformation of the sp2 structure to sp3 structure.

A large number of research works have focused on the mechanical properties of graphene-based epoxy nanocomposites [65-79]. High levels of reinforcement have been achieved by Cha *et al*. through non-covalent functionalization [67] of graphene nanoplatelets (GNPs) by melamine. The addition of 2 wt% functionalized GNPs led to an increase of the Young's modulus, tensile strength and fracture toughness of the nanocomposites. The reinforcing efficiency of GNPs was higher than their CNTs counterparts, compared to either pristine or functionalized CNTs. This lies in the fact that 2D GNPs are able to create a larger polymer/filler interface compared to 1D CNTs, which leads to an improved stress transfer efficiency, and also prevents the nanofiller pull out or debonding under strain. The functionalization process ultimately led to an improved dispersion of the samples by preventing agglomeration and restacking of the fillers and promoting interactions between the nanofillers and the epoxy. It should be mentioned at this point that a number of works have reported a decrease of the tensile strength of the epoxy matrix as a result of the introduction of graphene-related materials, even at very low filler contents (~0.01 vol%) [68, 69]. Tensile strength is highly sensitive to the presence of defects, which concentrate stress and act as failure points during the application of strain. Therefore, the introduction of 2D fillers which contain a high density of defects within a relatively high-strength polymer (such as epoxy resins) will not effectively increase the tensile strength. Additionally, strong interfacial adhesion needs to be ensured in order to form bonds between the matrix and the filler, allowing efficient stress transfer hence to achieve high tensile strength; for this purpose, 2D materials with functional groups on their surface are more suitable for maximisation of reinforcement.

Two-dimensional BN nanosheets (BNNSs) have also been used to mechanically reinforce epoxy nanocomposites [80-86]. As mentioned earlier, even though the elastic modulus of hBN is slightly lower compared to graphene, the increase in layer number (up to nine layers) is not expected to degrade significantly its intrinsic mechanical properties [28]. Lin *et al*. [81] prepared epoxy/hBN (20 wt.%) nanocomposites using solution blending and used magnetic field to align the hBN nanoplatelets in the epoxy matrix. As expected, vertical alignment (in the z direction) improved the elastic modulus of the composites significantly. In another study, non-covalent functionalization via 1-pyrenebutyric acid (PBA) was used to toughen BNNS/epoxy nanocomposites [82]. The fracture toughness of the neat epoxy was doubled with only 0.3 wt% PBA-BNNSs. The reinforcing mechanism is similar with the case of GNPs described earlier, where the non-covalent functionalization prevented BNNSs from aggregation and increased the contact area between BNNSs and the epoxy matrix.

Other 2D materials such as TMDCs [87-92], MXenes [93-96] and BP [97, 98] have also been used as mechanical reinforcing agents in epoxy nanocomposites. For example, Eksik *et al*. [87]

exfoliated $MoS_2$ powders in 1-vinyl-2 pyrrolidone assisted by sonication and prepared $MoS_2$/epoxy nanocomposites by solution blending. It was found that even at low contents (0.2 wt.%), $MoS_2$ can improve the mechanical properties of epoxy resins effectively. The Young's modulus and fracture toughness were slightly improved, while a more pronounced increase was recorded for the tensile strength. The enhancements were lower than the ones from their graphene counterparts; however, $MoS_2$ provides an alternative when the mechanical reinforcement of electrically insulating polymer composites is targeted.

Indicative literature works on the mechanical properties of epoxy nanocomposites with 2D materials can be seen in **Table 3**. Overall, graphene is among the most effective mechanical reinforcing agent in epoxy nanocomposites. Other 2D materials such as hBN and MXenes are also suitable for mechanical reinforcement considering their good mechanical properties and strong interlayer interactions. Furthermore, various 2D nanofillers can work synergistically with other nanofillers (2D-2D, 2D-1D, and 2D-0D) to further improve the mechanical properties of epoxy nanocomposites [99]. For example, 2D-2D hybrids can be used to construct ordered and laminated structures to achieve synergistic effects while the combination of 2D and 1D fillers can fabricate a 3D structure which is able promote the dispersion of each filler within a polymer matrix. Additionally, 0D fillers can be used to decorate the surface of 2D nanosheets to improve dispersion, properties and prevent agglomeration of 2D nanosheets [99]. The hybridisation process can also enhance the matrix-filler interactions, depending on the functionalisation route, while it can also counterbalance some of the disadvantages of a 2D filler. As a result, the use of hybrid fillers can lead to a high-performing nanocomposite with a diverse set of properties.

**Table 3.** Mechanical properties of epoxy nanocomposites reinforced with 2D materials. When two or three different optimum filler fractions are reported, each fraction corresponds to the maximum modulus, strength and toughness, respectively.

| Filler | Optimum filler fraction | Tensile modulus (GPa)/ Increase (%) | Tensile strength (MPa)/ Increase (%) | Fracture toughness (MPa m$^{1/2}$)/ Increase (%) | Ref. |
|---|---|---|---|---|---|
| GNP | 4 wt% | 2.1/21 | – | – | [65] |
| GNP | 0.1 wt% | 2.8/31 | 55/40 | 1.0/53 | [66] |
| GNP | 5 wt% | 3.2/18 | 95/1 | - | [68] |
| f-GNP | 0.5 wt% | –/15 | –/38 | –/82 | [70] |
| f-GNP | 2 wt% | 3.3/71 | 86/23 | 1.0/124 | [67] |
| f-GNP | 0.6 wt% | 2.7/33 | 48.2/99 | 0.8/232 | [71] |
| GO | 1 wt% | 2/42 | – | – | [72] |
| GO | 0.1, 0.5, 1 wt% | 3/12 | 65/13 | 0.7/63 | [73] |
| GO | 0.2 wt% | – | 50/121 | – | [74] |
| f-GO | 0.25 wt% | 3.1/13 | 53/75 | 0.5/41 | [75] |
| rGO | 0.3 wt% | 1.8/47 | 46.5/47 | – | [76] |
| rGO | 0.2 wt% | - | - | 0.5/52% | [77] |
| rGO | 1 wt% | – | – | 0.7/314 | [78] |
| GNP/CNT | 0.5 wt% | 2.2/40 | 50/36 | – | [79] |
| BN | 0.4 wt% | – | 65.6/118 | – | [85] |
| f-BN | 20 wt% | 2.7/68 | – | – | [81] |
| f-BN | 0.3 wt% | 2.7/21 | 46.7/54 | 0.75/107 | [82] |
| f-BN | 1 vol% | 2.9/66 | – | – | [86] |
| f-BN | 10 wt% | – | 25/100 | – | [83] |
| BN/MWCNT | 0.5/0.3 wt% | 3.2/38 | 60/25 | – | [84] |
| MoS$_2$ | 0.2 wt% | 3.4/9 | 68/33 | 1.0/66 | [87] |
| f-MoS$_2$ | 0.2 wt% | 0.9/26 | 39.9/23 | – | [88] |
| f-WS$_2$ | 0.25 wt% | – | – | 0.94/83 | [89] |
| MoS$_2$/MWCNT | 1 wt% | 2.6/47 | 45/49.6 | – | [90] |
| MoS$_2$/CF | 0.8 wt% | 1.2/53 | 31.26/77 | – | [91] |
| MoS$_2$/SiO$_2$ | 3 wt% | 1.2/38 | 32.32/81 | – | [92] |

| Ti₃CN | 40 wt% | 4.5/93 | – | – | [93] |
|---|---|---|---|---|---|
| Ti₃CN | 90 wt% | 4.5/182 | – | – | [93] |
| Ti₃C₂Tₓ | 5 wt% | 3.6/21 | – | – | [94] |
| Ti₃C₂Tₓ | 1.2 wt% | – | 53/25 | – | [95] |
| Ti₃C₂Tₓ | 0.2 wt% | 2.6/35 | 70.5/51 | 0.74/45 | [96] |
| f-BP | 3 wt% | – | 73.5/21 | – | [97] |

ᵃ f-: functionalized; CF: carbon fibre.

## 4.2 Tribological properties

Friction is universal and inevitable in our life, and it often causes the waste of energy and resources; therefore, it is important to monitor and control friction and wear. Benefiting from their low cost, easy processing and low friction, epoxy resins have been used in components that are commonly subjected to wear. For example, epoxy resins have been widely applied in the construction industry for coating on floors, counters, bar-tops and much more. The use of epoxy resins can generate a smooth, glossy and hard-wearing surface. However, the load carrying capacity and thermal properties of epoxies are considerably lower than metals and ceramics. Their 3D crosslinked network structure attributes poor tribological properties and low wear resistance. As a result, nanoreinforcements have been used to improve these properties. Various 2D materials such as graphene and $MoS_2$ exhibit low friction coefficients owing to their easy shear capability on their atomically smooth surfaces [100]. Additionally, the high strength of 2D materials makes them resistant to wear while their nanoscale dimensions lead to a high surface to volume ratio compared to conventional composites. Finally, the impermeability of 2D materials to a number of gases and liquids can delay the corrosive and oxidative processes. These characteristics make 2D materials effective in enhancing the tribological performance of epoxy nanocomposites. The friction coefficient and wear rate are the two main parameters that describe the tribological performance of polymer composites, and lower values represent better tribological performance. The friction coefficient is the ratio of the frictional force to normal force, and the wear rate, $W_s$, can be calculated by the following equation [101]: $W_s = \frac{\Delta m}{\rho L F_N}$ where $\Delta m$ and $\rho$ are the mass loss and density of the composite, $L$ is the sliding distance and $F_N$ is the normal force.

Graphene [102-105] and MoS$_2$ [106-110] have been commonly used as solid lubricants to enhance the tribological performance of epoxy nanocomposites. Chen *et al*. [106] studied the effects of GO, CNT and MoS$_2$ (at 1.2 wt%) on the tribological performance of epoxy nanocomposites under dry sliding. The increase in sliding velocity or applied load increases the friction coefficient and wear rate, as high speeds or loads increase temperature during friction and this results in decreased adhesion and increased wear. The addition of CNTs, GO, MoS$_2$ or their hybrids reduced the friction coefficient and wear rate of the epoxy nanocomposites effectively (**Figure 4a and 4b**). Compared to CNTs and GO, the sample filled with neat MoS$_2$ displayed a higher reduction in friction due to its excellent self-lubricating properties. The epoxy nanocomposites with the hybrid filler (a combination of all three fillers) exhibited the best friction performance and the coefficient of friction of the epoxy was decreased by more than 90% (**Figure 4a**). Similarly, the wear rate of the pure epoxy was reduced by more than 95% with the hybrid CNTs, GO and MoS$_2$ filler, as shown in **Figure 4b**. The tribology mechanisms are explained schematically in **Figure 4c and 4d**. The neat epoxy is not resistant to wear and can be easily worn off with the formation of spalling pits and large debris. For epoxy nanocomposites, GO and CNTs with their high mechanical strength, act as load bearers at surface valleys to prevent the matrix from being further worn and MoS$_2$ acts as a self-lubricating layer to reduce the friction coefficient. In addition, a transfer film is formed on the steel ball, which can reduce the friction coefficient and wear rate by the formation of a low-strength conjunction at the interface.

Overall, graphene and MoS₂ are both excellent lubricating additives in epoxy nanocomposites, and they can function additively to further improve the tribological performance of epoxy nanocomposites.

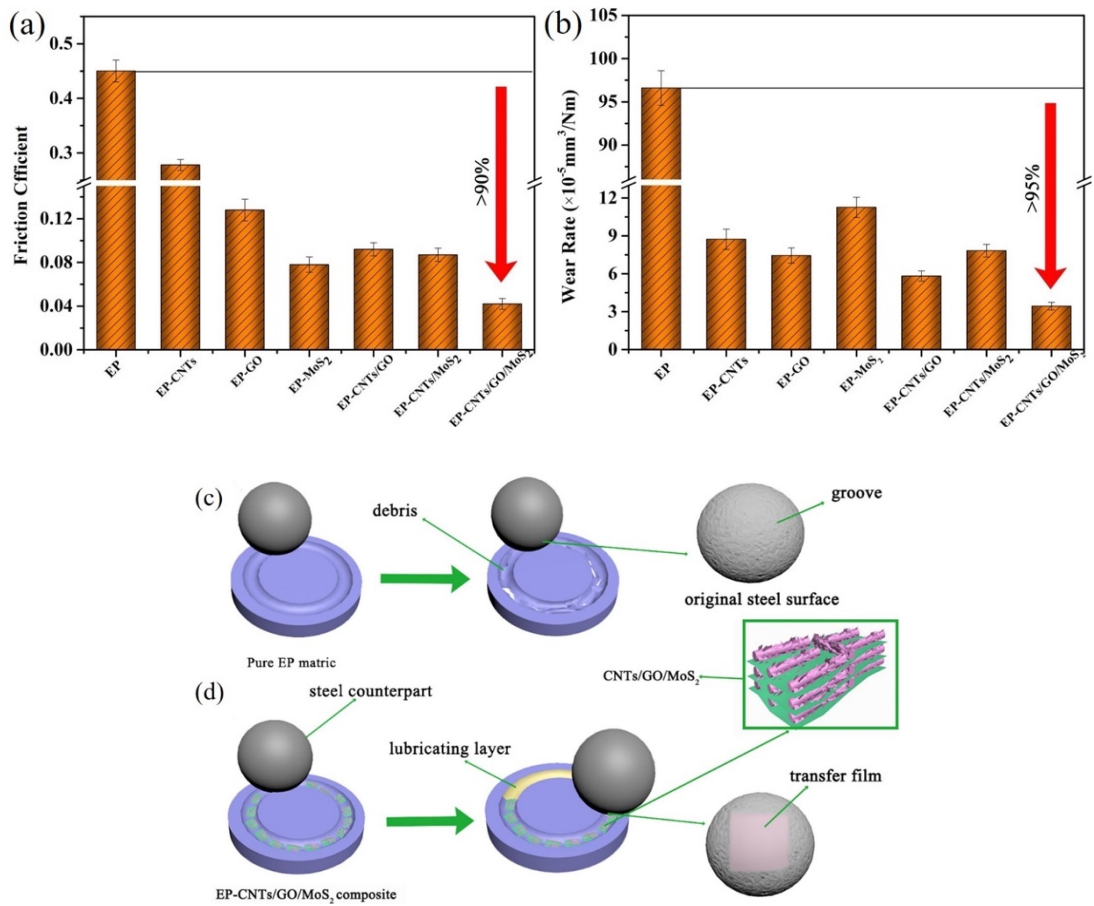

**Figure 4**. (a, b) Variations of friction coefficient and wear rate of epoxy nanocomposites with CNTs, GO and MoS₂. (c, d) Tribology mechanisms of pure epoxy matrix and epoxy/CNTs/GO/MoS₂ composite. The pure epoxy matrix can be easily worn off and large debris and groove are produced during the wearing process. For the hybrid epoxy nanocomposites, a lubricating layer and transfer film are produced during the tests that lead to a reduction of the wear rate. Reproduced with permission from [106]. Copyright 2018 Elsevier.

Apart from graphene and MoS₂, the effects of other 2D materials such as hBN [111-114] and MXenes [115-118] on the tribological performance of epoxy nanocomposites have also been investigated. Yu *el al.* [112] studied the effects of functionalized cubic boron nitride (FC-BN) and functionalized hexagonal boron nitride (FH-BN) on the tribological performance of epoxy nanocomposites. On the one hand, the BN nanosheets exhibit poorer self-lubricating properties compared to graphene and $MoS_2$ due to stronger interlayer interactions; thus, the friction coefficient of the epoxy matrix decreased only slightly with the addition of BN. On the other hand, the high strength of BN nanosheets increased the hardness of the composites and prevented the propagation of cracks; therefore, the wear rate of the epoxy matrix was sharply reduced with the addition of 0.5 wt% FH-BN. Recently, Meng *et al.* [115] studied the tribological performance of epoxy nanocomposites reinforced with $Ti_3C_2$ MXene. The coefficient of friction of the pure epoxy decreased by 76.3% with the introduction of 1 wt% $Ti_3C_2$, and the wear rate decreased by 67.4% with 3 wt% $Ti_3C_2$. The distribution of $Ti_3C_2$ in the epoxy was homogeneous due to the well-ordered 3D structure. During the friction process, the cracks propagate into the aligned nanosheets and form flake debris. Some of the debris can slide to reduce the friction coefficient and wear rate. Furthermore, few-layer $Ti_3C_2$ can shear under force, which can lead to efficient friction reduction. The hardness of $Ti_3C_2$ is responsible for the reduction in wear rate.

The tribological performance of various epoxy nanocomposites reinforced with 2D materials is summarized in **Table 4**. Graphene and $MoS_2$ are more effective in the reduction of friction coefficient due to their self-lubricating properties, while graphene, hBN and MXenes are effective towards the reduction of wear rate due to their high strength. Overall, the combination of high mechanical strength and excellent lubricating properties makes 2D materials suitable for the reduction of the friction coefficient and wear rate of epoxy resins. These 2D materials can also work additively or synergistically to further improve the tribological performance of epoxy nanocomposites. For example, a 99% reduction in friction coefficient of an epoxy has been achieved with the incorporation of a hybrid graphene-$MoS_2$ filler [104].

**Table 4.** Tribological properties of epoxy nanocomposites reinforced with 2D materials.

| Filler | Optimum filler fraction | Operating conditions | Coefficient of friction/ Decrease (%) | Wear rate ($10^{-6}$ mm$^3$/Nm)/ Decrease (%) | Ref. |
|---|---|---|---|---|---|
| GO | 0.5 wt% | 0.5 MPa, 1 m/s | 0.9/-1 | 60/80 | [102] |
| GO | 0.5 wt% | 1 MPa, 1 m/s | 0.47/-100 | 130/87 | [102] |
| GONR | 0.6 wt% | 100 N, 400 rpm/min | 0.6/68 | – | [103] |
| Gr/$MoS_2$ | 5 wt% | 1 N, 0.41 m/s | 0.25/99 | – | [104] |
| f-rGO/PTFE | 1/10 wt% | 5N, 4.2 Hz | 1.12/88 | –/21 | [105] |
| $MoS_2$ | 10 wt% | 15N, – | 0.53/29 | 607/-6 | [107] |
| $MoS_2$ | 6 wt% | 5 N ,5 mm/s | 0.63/65 | – | [108] |
| $MoS_2$/GO/CNT | 1.25 wt% | 4 N, 200 rpm/min | 0.45/90 | 970/95 | [106] |
| $MoS_2$/CF | 1.25 wt% | 4 N, 200 rpm/min | 0.42/82 | 689/88 | [109] |
| $MoS_2$/hBN | 1.5 wt% | 4 N, 200 rpm/min | 0.4/80 | 675/88 | [110] |
| BN | 0.5 wt% | 10N, 1.5 m/s | 0.72/54 | 120/67 | [113] |
| hBN | 0.5 wt% | 5N, 2 Hz | – | 225/33 | [111] |
| f-hBN | 0.5 wt% | 5 N, 2 Hz | – | 3500/43 | [114] |
| f-hBN | 0.5 wt% | 5 N, 5 Hz | 0.65/12 | 2228/73 | [112] |
| $Ti_2CT_x$ | 2 wt% | 98 N, 0.3 m/s | 0.67/66 | – | [117] |
| $Ti_3C_2$ | 3.7 wt% | 5 N, 2 Hz | 0.71/76 | 304/67.4 | [115] |
| f-$Ti_3C_2T_x$ | 0.5 wt% | 3 N, 2 Hz | 0.54/34 | 711/73 | [116] |
| $Ti_3C_2$/Gr | 0.25/0.25 wt% | 5 N, 2 Hz | 0.58/9.8 | 1081/89 | [118] |

[a] f-: functionalized; GONR: graphene oxide nanoribbon; PTFE: polytetrafluoroethylene;

### 4.3 Thermal conductivity

Epoxy resins gain huge popularity in electronics industry, due to their ability to offer protection against chemicals and high temperature. The use of epoxy resins allows protection of electronic components against dust, moisture and short circuits. The rapid development in electronics and e-vehicles requires higher power components with reduced size, which means more heat being released. This drives the synthesis of resins with high thermal conductivity to improve heat dissipation from electrical components. However, pure epoxy resins display low thermal conductivity (~0.2 Wm$^{-1}$K$^{-1}$), which is not able to meet the market requirements. The incorporation of thermally conductive fillers is a routine method to improve their performance. Among the list of 2D materials, graphene and hBN display high thermal conductivities and therefore hold the best potential to improve the thermal properties of epoxy resins [119]. The thermal conductivities of other 2D materials are relatively low compared to graphene and hBN, which can restrain their effectiveness in thermal reinforcement [119]. Additionally, 2D materials can be used to construct hierarchically ordered conductive networks (also in combination with other fillers), which can lead to increased conductivities in the final epoxy nanocomposites. The thermally conductive mechanism in polymer nanocomposites is shown in **Figure 5a and 5b** [120]. When the fillers are dispersed in a matrix without inter-filler network (**Figure 5a**), the composite thermal conductivity is described by the series model, $k_c = \left[\frac{V_m}{k_m} + \frac{V_f}{k_f}\right]^{-1}$, where $k_c$, $k_m$ and $k_f$ are the thermal conductivities of the composites, matrix, and filler, respectively [121]. When fillers increase to a critical loading and the inter-filler network is formed in a matrix (**Figure 5b**), the composite thermal conductivity is given by the parallel model, $k_c = V_m k_m + V_f k_f$ [121]. These two models set up the lower bound and upper bound for most of the experimental results.

In the work of Chinkanjanarot *et al.* [122], a multiscale modelling approach was proposed using molecular dynamics and micromechanics to evaluate the thermal conductivity of a cycloaliphatic epoxy reinforced with GNPs. The results revealed that the functionalisation of GNPs leads to better dispersion within the matrix, which in turn leads to increased composite thermal conductivity, compared to their unfunctionalised counterparts. It should be noted that the simulations included perfectly flat and defect-free graphene materials with a thickness up to 4 layers, which is hard to achieve or use in bulk quantities, while the orientation of the fillers was random.

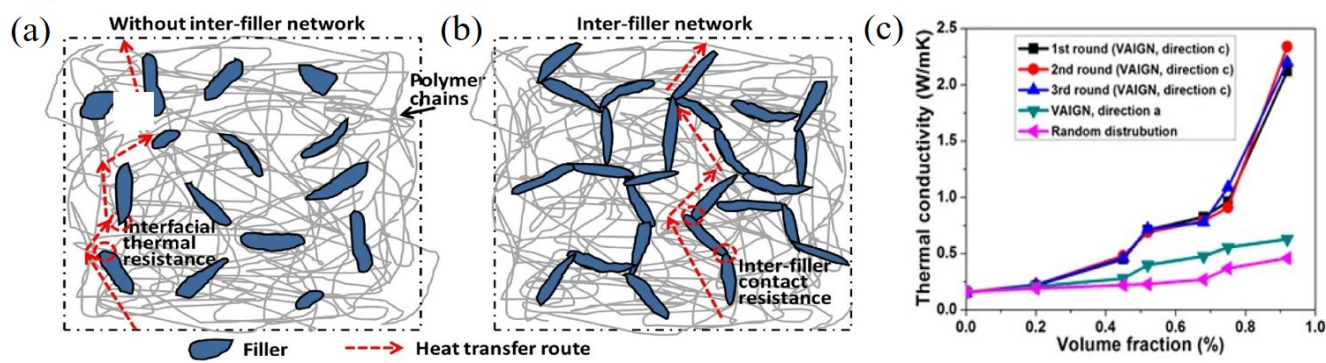

**Figure 5.** (a, b) Heat transfer in polymer nanocomposites without inter-filler network and with inter-filler network. Reproduced with permission from [120]. Copyright 2018 Elsevier. (c) Variations of thermal conductivity with graphene volume fraction. Reproduced with permission from [123]. Copyright 2016 American Chemical Society.

Graphene has been widely used to improve the thermal conductivity of epoxy resins [51, 52, 123-133]. As mentioned earlier, it is important to form percolating networks through the construction of 3D interconnected structures or the increase of filler loadings to deliver high thermal conductivities. Lian *et al*. [123] improved the thermal conductivity of an epoxy resin by more than one order of magnitude with only 0.92 vol% graphene, due to the formation of 3D rGO networks obtained by freeze-drying. This value is much higher compared to the samples where rGO was randomly oriented (at the same loadings) (**Figure 5c**). The formation of the 3D network due to freeze-drying played a vital role in the improvement of thermal conductivity, as heat was efficiently transferred through the 2D nanosheets. Exceptional improvements of thermal conductivity have been also reported by the Regev group [51, 52], again highlighting the importance of the creation of an oriented filler network in the composite. In their highly-conductive graphene-epoxy nanocomposites, compression forces were applied either by zirconia balls during dispersion that resulted in gap closure and network formation of GNPs [51] or by hot pressing the nanocomposite before curing, to achieve minimisation of the filler-filler distance (and easier phonon conduction) while reducing the air voids in the composites [52]. Both series of composites exhibited a rapid thermal conductivity increase with filler loading, yielding very high ultimate values, at large filler contents (~40 wt%).

The excellent thermal properties of graphene have also been utilised in epoxy nanocomposites to perform a resistive electrothermal/heating process (otherwise known as Joule heating) for the curing of the epoxy resin [134, 135]. Xia *et al*. demonstrated that the dispersed GNPs (10 wt.%) in epoxy resins can convert the applied electric voltage into thermal energy and successfully cure the composites via Joule heating [135]. Excellent electrical and mechanical properties with reduced microvoids were obtained by this novel method compared to traditional oven curing, thanks to the outstanding electrothermal properties of graphene and its preferential orientation within the matrix.

Boron nitride nanosheets (BNNSs) have also been widely used to improve the thermal conductivity of epoxy nanocomposites [50, 136-142]. Even though the thermal conductivity of BN is lower compared to that of graphene, high-performing, thermally conductive epoxy nanocomposites have been fabricated with high contents of BNNSs [50, 136, 137]. As mentioned earlier, the fabrication of aligned 2D or 3D networks is one of the most effective ways to transfer heat within a nanocomposite. Han *et al*. [136] prepared BNNS/epoxy nanocomposites by using a polydimethylsiloxane (PDMS) wedge to generate temperature gradients and guide ice crystals into a lamellar morphology followed by the alignment of BNNSs into the same structure. Then, nacre-like composites were obtained by freeze-drying and infiltration of the epoxy resin. Additional, randomly distributed and uniaxially aligned composites were prepared for comparison reasons (**Figure 6a-c**). The thermal conductivities of the fabricated composites at different orientations and with different filler loadings are shown in **Figure 6d and 6e** respectively. It can be seen that the thermal conductivity increases significantly as a result of the long-range lamellar orientation (**Figure 6d**). In addition, from measurements in both the parallel ($\lambda_{\parallel}$) and perpendicular ($\lambda_{\perp}$) directions with respect to the lamellar layers, the increase of conductivity for BNNS loading up to 20 vol% BNNS in the $\lambda_{\parallel}$ direction is almost one order of magnitude higher compared to the $\lambda_{\perp}$ direction (**Figure 6e**).

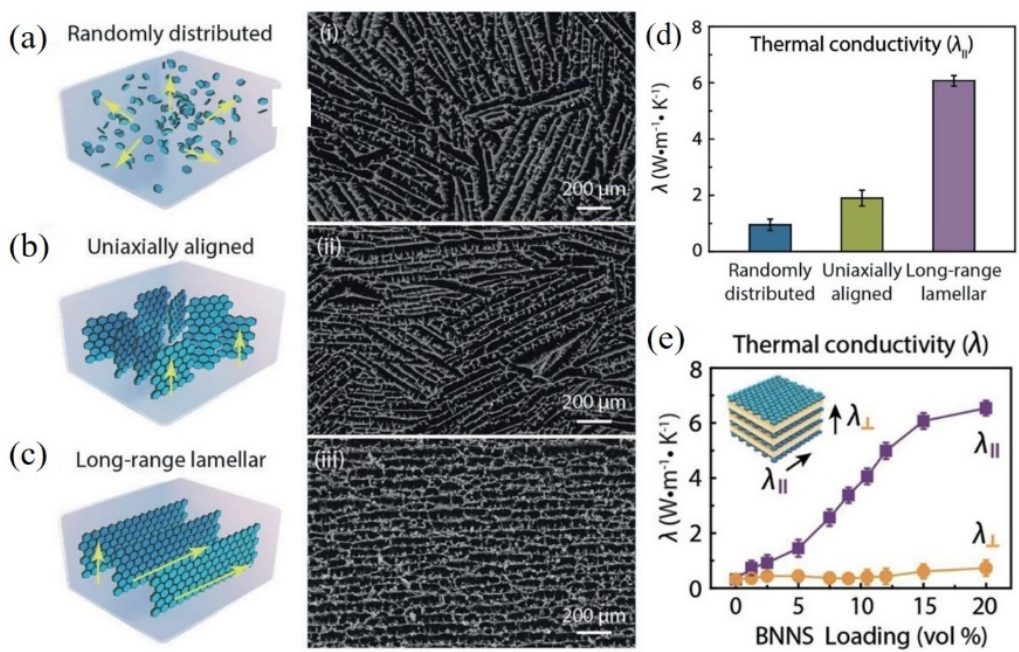

**Figure 6**. (a-c) Randomly distributed, uniaxially aligned and long-range lamellar network of BNNS/epoxy nanocomposites. (i-iii) SEM images of the network corresponding to (a), (b) and (c). (d) Thermal conductivities of the nanocomposites with different filler network. (e) Variations of thermal conductivity of BNNS/epoxy nanocomposites with filler loading for parallel and perpendicular lamellar layers between epoxy resin layers. Reproduced with permission from [136]. Copyright 2019 Wiley-VCH Verlag GmbH & Co. KGaA.

Recently, MXene/epoxy nanocomposites received attention as a result of their thermal management properties [143-146]. For example, Ji *el al*. [145] fabricated hybrid MXene nanosheets/Ag nanoparticles, which can act as 3D heat transfer channels for epoxy nanocomposites, using the ice template method. The decorated Ag nanoparticles on the surface of the MXene nanosheets helped reducing the contact resistance, while the vertically-aligned MXene nanosheets acted as heat conduction skeleton. Using this method, the in-plane and through-plane thermal conductivity of the composites were significantly improved.

The thermal conductivity of various epoxy nanocomposites reinforced with 2D materials is summarized in **Table 5**. Overall, the incorporation of 2D materials and especially graphene and BN can improve the thermal conductivity of epoxy resins significantly as a result of their excellent inherent thermally conductive properties and their large lateral dimensions that allow the creation of conductive pathways for an efficient phonon transport. It is extremely important to construct aligned 2D or 3D connected networks by using high aspect ratio nanoplatelets, in order to minimize phonon scattering, induced by the particle-particle interfaces, and thus improve the thermal conductivity effectively. Hybrid nanofillers can also be used to construct heat transfer channels to improve the thermal conductivity of epoxy nanocomposites as a result of the fabrication of a complex filler network and enhancement of the effective aspect ratio of a single filler [50-52, 127, 138, 139].

**Table 5.** Thermal conductivity of epoxy nanocomposites reinforced with 2D materials.

| Filler | Optimum filler fraction | Thermal conductivity ($Wm^{-1}K^{-1}$) | Thermal conductivity Increase (%) | Ref. |
|---|---|---|---|---|
| GNP | 3 wt% | 0.5 | 144.6 | [128] |
| GNP | 2.8 vol% | 1.5 | 650 | [125] |
| GNP | 24 vol% | 12.4 | 6800 | [51] |

| GNP | 45 vol% | 11 | 5400 | [129] |
|---|---|---|---|---|
| f-GNP | 20 wt% | 5.8 | 2800 | [132] |
| GNP/MWCNT | 0.9/0.1 wt% | 0.32 | 146.9 | [133] |
| Graphite/Graphene hybrid | 10 vol% | 5.1 | 2300 | [126] |
| Graphite/GNP hybrid | 70 wt% (40/30 wt%) | 16 | 7900 | [52] |
| Graphene web | 8.3 wt% | 8.8 | 4300 | [124] |
| rGO | 0.92 vol% | 2.1 | 1231 | [123] |
| f-rGO | 4 phr | 1.9 | 855 | [130] |
| GNP/MoS$_2$/CNT | 20 wt% | 4.6 | 2300 | [127] |
| GNP/Cu nanoparticles | 75 wt% (40/35 wt%) | 13.5 | 6650 | [131] |
| GNP/BN | 17 vol% (15/2 vol%) | 4.7 | 2250 | [51] |
| BN | 40 wt% | 6 | 2900 | [140] |
| BN | 45 vol% | 5.5 | 2650 | [129] |
| BN | 20 vol% | 6.5 | 3170 | [136] |
| BN | 9.3 vol% | 2.8 | 1681 | [141] |
| f-BN | 44 vol% | 9 | 4400 | [137] |
| BN/Ag | 20 wt% (10/10 wt%) | 1.1 | 465 | [138] |
| BN/Ag | 62.2 wt% | 23.1 | 11500 | [50] |
| BN/cellulose skeleton | 9.6 vol% | 3.1 | 1400 | [142] |
| hBN/MoS$_2$ | 1 wt% (0.5/0.5 wt%) | 0.9 | 203 | [139] |
| Ti$_3$C$_2$ | 1 wt% | 0.59 | 141.3 | [143] |
| Ti$_3$C$_2$T$_x$ | 30 wt% | 3.14 | 1470 | [144] |
| MXene/Ag | 15.1 vol% | 1.79 | 795 | [145] |
| Ti$_3$AlC$_2$/CF/cellulose | 30.2 wt% (20/10.2 wt%) | 9.7 | 4509 | [146] |
| f-BP | 3 wt% | 0.4 | 66 | [97] |

## 4.4 Electrical properties

The electrical insulating nature of epoxies has been utilised for potting and encapsulation in the electronics and electrical industries. However, electrically conductive resins and adhesives are necessary in some applications. Conductive resins and adhesives can be used as spraying or electrostatic painting on fuel lines, sensitive, high-temperature electronics and external components on vehicle or wind turbines to prevent sparks. An important application of electrically conductive epoxy nanocomposites is electromagnetic interference (EMI) shielding. High electrical conductivity is required for EMI shielding, due to the interactions between charge carriers and EM fields [147]. The incorporation of electrically conductive 2D materials can transform an epoxy from insulating to conductive upon the establishment of a percolated network [148]. According to the percolation theory, the electrical conductivity of composites can be described from the following relationship: $\sigma = \sigma_0(V_f - V_c)^t$ where $\sigma$ is the electrical conductivity of the composite, $\sigma_0$ is a preexponential factor, dependent on the electrical conductivity of the filler, $V_c$ is the percolation threshold in volume fraction, and $t$ is the conductivity exponent. The value of $t$ is around 1.1- 1.3 for 2D systems and 1.6-2.0 for 3D systems [148]. Therefore, the electrical conductivity of nanocomposites does not follow a linear relationship with the filler loading, as shown in **Figure 7a**. At low loadings, the polymer is insulating or displays low electrical conductivity. As the filler loading reaches the percolation threshold, a dramatic increase in conductivity can be observed due to the formation of a conductive network for electronic charge carriers (electrons or holes). The conductivity then reaches a plateau with less obvious increment in conductivity upon further increase in filler loading. Similar to thermal conductivity, it is of great importance to construct aligned 2D or 3D interconnected networks to impart electrical conductivity effectively. Additionally, fillers with large lateral dimensions, especially with larger aspect ratios, are favoured for the formation of percolating networks as they can reduce the percolation threshold.

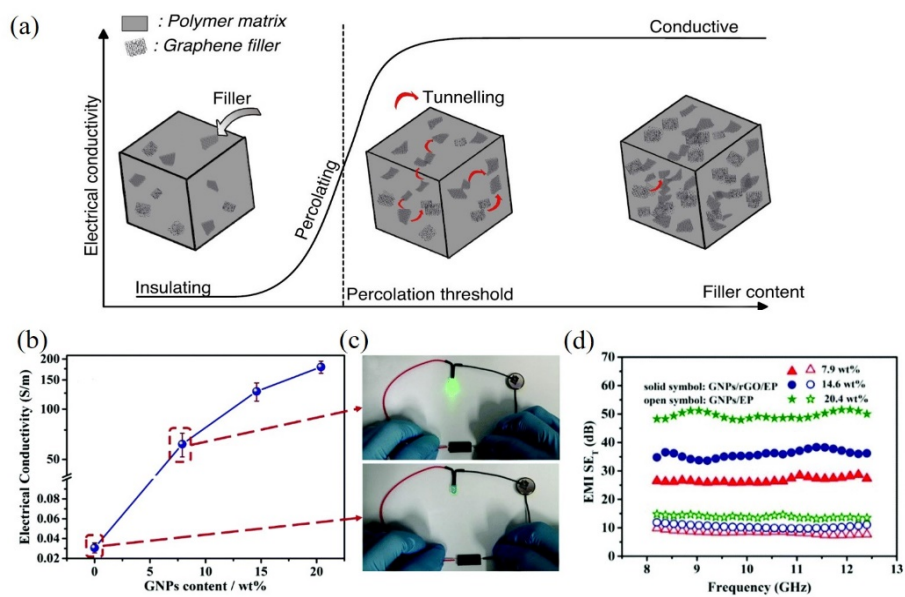

**Figure 7.** (a) A schematic of electrical conductivity mechanism in polymer nanocomposites with filler content. Reproduced with permission from [149]. Copyright 2018 IOP. (b) Electrical conductivity of the epoxy nanocomposites with various GNP contents. (c) LED lamp at 3 V using rGO/epoxy and GNPs/rGO/epoxy nanocomposites with 0.1 wt% rGO and 7.9 wt% GNPs as conductive elements. (d) Variations of total EMI SE ($SE_T$) of the epoxy nanocomposites with GNPs and rGO. Reproduced with permission from [150]. Copyright 2019 The Royal Society of Chemistry.

Considering its high electrical conductivity and large aspect ratio, graphene can impart a relatively low percolation threshold and can improve the electrical conductivity, as well as the EMI shielding effectiveness (SE) of epoxy resins effectively [150-156]. Recently, epoxy nanocomposites with high electrical conductivity and high EMI SE reinforced by both GNPs and rGO were manufactured through the construction of 3D conductive networks [150]. The variations of electrical conductivity and EMI SE of the composites as a function of the filler loading are shown in **Figure 7b and 7d**. The electrical conductivity of rGO/epoxy nanocomposites is low, due to the presence of defects in the rGO structure; however, this value increased with the addition of 7.9 wt% GNPs due to the intrinsic electrical conductivity of GNPs. The effectiveness of the 3D hybrid filler network on electrical conductivity and EMI SE can be observed in **Figure 7d**, where the total EMI SE ($SE_T$) of epoxy nanocomposites prepared by freeze-drying (solid symbols) is much higher compared to their counterparts prepared by solution blending (open symbols). The epoxy nanocomposites with 0.1 wt% rGO and 20.4 wt% GNPs exhibit the highest EMI SE in the X-band range as a result of the conductive networks, which act as dissipating mobile charge carriers and interact with the incident electromagnetic waves. Moreover, the incoming electromagnetic waves are multiply absorbed, reflected and scattered by the 3D structure. Overall, not only the intrinsic properties of 2D fillers, but also the formation of 3D interconnected networks contribute to the enhancement of electrical conductivity and the EMI SE of epoxy nanocomposites.

Apart from graphene, MXenes also exhibit high electrical conductivity as well as excellent EMI SE and have been used in epoxy nanocomposites [94, 157-159]. Wang *et al*. [94] studied the effects of annealing on the electrical properties of $Ti_2C_3T_x$/epoxy nanocomposites. Epoxy nanocomposites with annealed $Ti_2C_3T_x$ displayed higher electrical conductivity compared to non-annealed epoxy-$Ti_2C_3T_x$ nanocomposites. The epoxy nanocomposites with 15 wt% annealed $Ti_2C_3T_x$ exhibited the highest EMI SE as a result of the removal of the surface functional groups from $Ti_2C_3T_x$, which means that more dipoles were formed and the electron transport capability was enhanced.

Another effective strategy to improve the electrical and EMI shielding performance of epoxy nanocomposites is the fabrication of conductive networks with a hybrid graphene-MXene filler. As seen from **Figure 8a**, Song *et al*. [158] prepared epoxy nanocomposites reinforced with both rGO and MXene nanoplatelets. The electrical conductivity and EMI SE of the epoxy nanocomposites are shown in **Figure 8b** and **8c**. As it can be seen, the electrical conductivity and the EMI SE increased with the increase of MXene loading, while the epoxy nanocomposite with maximum loading of both fillers displayed the highest electrical conductivity and the highest EMI SE.

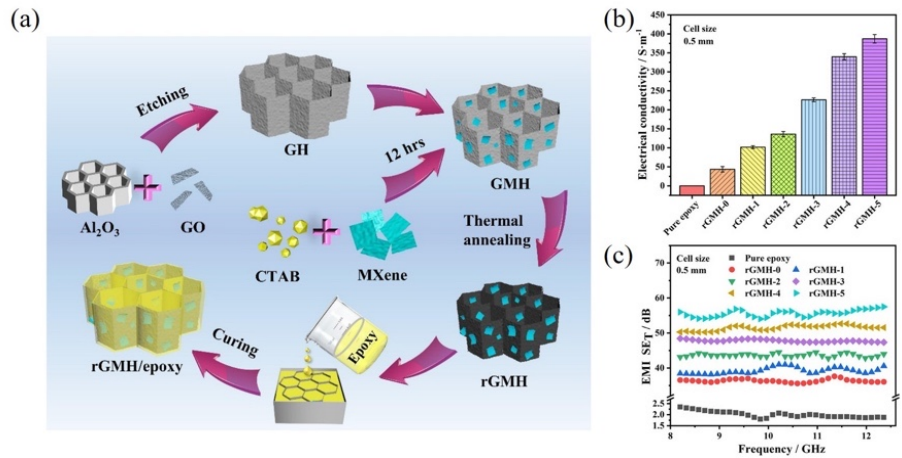

**Figure 8**. (a) Preparation of honeycomb structural epoxy nanocomposites with rGO and MXene. (b, c) Variations of electrical conductivity and EMI SE of the epoxy nanocomposites with filler loading, where rGMH represents honeycomb structural rGO-MXene. Reproduced with permission from [158]. Copyright 2020 Elsevier.

The electrically conductive graphene network in epoxy resins can also enable other functional properties such as electrical resistance-based sensing. Similar to strain and damage sensing based on percolated carbon nanotube network, the percolated graphene network can be utilised to monitor mechanical deformation as well as damage initiation and propagation in nanocomposites, providing feasible routes towards online health monitoring in epoxy-based structural applications or coatings. Kernin *et al.* demonstrated that rGO/epoxy nanocomposites with 0.5 wt.% filler loading show excellent cyclic sensing performance, with a gauge factor (GF) over 40 (**Figure 9a**) [53]. The change of distances and contacts between conductive fillers within the network during deformation can induce a measurable electrical sensing signal (**Figure 9b**). By correlating the sensing signals to strain, the deformation or damage of nanocomposite can be monitored in real time. Different loading conditions such as tensile and flexural, together with various failure modes in epoxy nanocomposites can be monitored via this method [160].

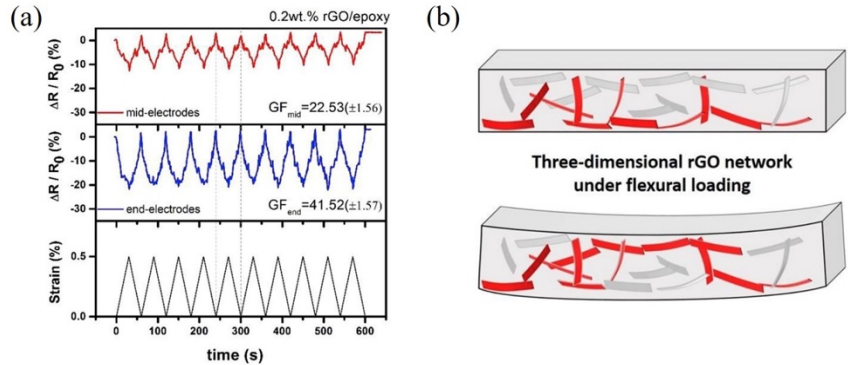

**Figure 9.** (a) rGO/epoxy nanocomposites under cyclic loading with strains up to 0.5%, showing high sensitivity (GF = 41.5) and a clear change in electrical resistance for each cycle, (b) schematic illustration of the side-view of the internal 3D network elucidating the reduced resistance upon flexural loading due to reduced inter-particle distances between rGO in compression. Reproduced with permission from [53]. Copyright 2019 Elsevier.

The electrical conductivities and the EMI SE of epoxy nanocomposites reinforced with graphene, MXenes and some hybrid fillers are summarized in **Table 6**. It can be seen that both 2D fillers are effective towards the electrical reinforcement of epoxy nanocomposites, while hybrid fillers can further improve the electrical performance of epoxy nanocomposites. A combination of 2D fillers with other conductive fillers (such as MWCNTs) can also aid towards improvement of the electrical conductivity and EMI SE of epoxy nanocomposites. Microscopically, the homogeneous dispersion and formation of a seamless, interconnected network of fillers can act effectively towards electrical percolation.

**Table 6.** Electrical conductivity and EMI SE of epoxy nanocomposites reinforced with 2D materials.

| Filler | Optimum filler fraction | Thickness (mm) | Electrical conductivity (S/m) | EMI SE (dB) | Ref. |
|---|---|---|---|---|---|
| Gr | 15 wt% | – | – | 21 | [154] |
| rGO | 1.2 wt% | 0.5 | 40.2 | 38 | [151] |
| rGO/CF | 0.5 wt% | 6 | 7.2 | 37.6 | [152] |
| GA/MWCNT/PANI | 1.2/0/83/2.58 wt% | 3 | 52.1 | 42 | [155] |
| rGO/Ag | 0.44/0.94 vol% | 3 | 45.3 | 58 | [156] |
| GO/f-Fe$_3$O$_4$ | 1.2/1/5 wt% | 3 | 27.5 | 35 | [153] |
| GNP/rGO | 20.4/0.1 wt% | 3 | 179.2 | 51 | [150] |
| Ti$_3$C$_2$T$_x$ | 1.2 wt% | – | $4.52 \times 10^{-4}$ | – | [95] |
| Ti$_3$C$_2$T$_x$ | 15 wt% | 2 | 105 | 41 | [94] |
| Ti$_3$C$_2$T$_x$ | 0.4 vol% | 2 | – | 34.5 | [157] |
| MXene/rGO | 3.3/1.2 wt% | 0.5 | 387.1 | 55 | [158] |
| MXene/rGO | 3.3/1.2 wt% | 1 | 105.8 | 35 | [158] |
| MXene/rGO | 3.3/1.2 wt% | 2 | 85.1 | 28 | [158] |
| MCF | 1.64/2.61 wt% | 2 | 184 | 46 | [159] |

[a] GA: graphene aerogel; PANI: polyaniline; MCF, Ti$_3$C$_2$T$_x$ MXene/C hybrid foam.

## 4.5 Flame retardancy

As mentioned before and after, epoxy resins have been extensively used as laminates and structural matrix materials, surface coatings, structural adhesives and electronic materials. However, untreated epoxy resins typically exhibit high flammability and high smoke production during combustion, which restricts applications requiring good flame resistance. As a result, the introduction of epoxies in a number of advanced applications is highly associated with the fabrication of materials that display reduced flammability. In flame retardant materials it is difficult to quantify the contributions from various elements of a system towards one dominant mechanism; however, the effective mechanism of flame retardancy can be divided into two actions: the gas-phase and the condensed-phase actions, as illustrated in **Figure 10** [161]. A number of different areas are involved in the flame retardancy mechanism when flame is directed to a polymer: the flame zone, the char layer, the thermal decomposition (or molten polymer) and underlying polymer zones. The char layer is critical as it controls heat and mass transfer between condensed and gas phases, while volatiles are generated in the thermal decomposition zone and then mitigate towards the flame zone. The condensed phase mechanism involves the formation of char layer on the surface which controls the heat and mass transfer between the gas phase and the condensed phase by blocking flammable volatiles from diffusion and shielding the polymer from air and heat. The gas phase mechanism includes the interruption of combustion process by radical absorption, as halogen-containing compounds can form less reactive halogen atoms by the reaction between specific radicals and highly reactive species. Different strategies such as chemical functionalization and incorporation of flame retardant additives have been used to improve the flame retardancy of epoxy resins. 2D materials have been studied as flame retardant additives, since they can facilitate the formation of a protective char layer in the condensed phase [162]. Additionally, 2D materials can also inhibit the combustion process, during the gas phase mechanism.

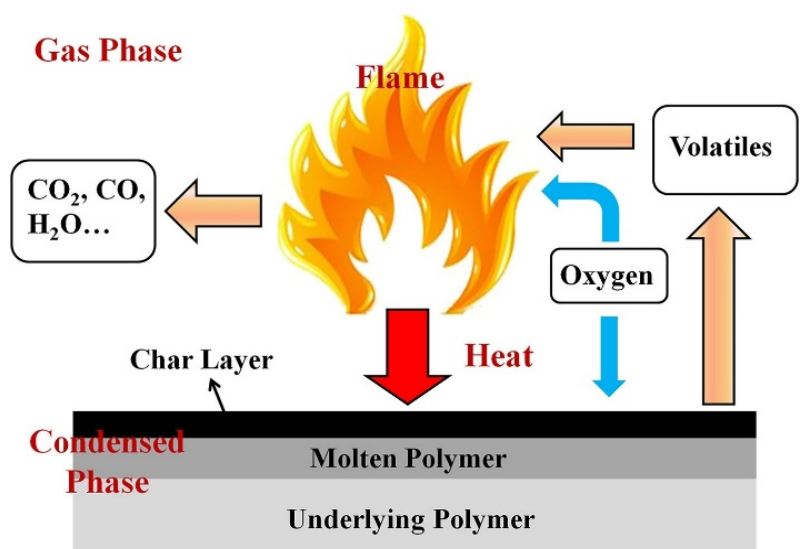

**Figure 10.** Flame retardant mechanisms in polymer nanocomposites. Reproduced with permission from [161]. Copyright 2017 Elsevier.

Graphene can be classified as a halogen-free flame retardant that can work as physical barrier and facilitate the formation of a char layer. It has been shown that graphene exhibits high flame resistance under natural gas flame and can be used to slow down the combustion of epoxy nanocomposites [163]. On the other hand, GO is highly flammable due to the effects of potassium salts (eg. $KMnO_4$ and $K_2S_2O_8$) that are used during its synthesis, which must be removed from GO in order to be considered for flame retardant applications [164]. Different methods such as the incorporation of flame retardant elements [165], polymer grafting [166] and the combination with other fillers [167, 168] have been adopted to improve the flame retardancy of epoxy nanocomposites with graphene and graphene derivatives. Feng *et al*. [165] studied the effects of phosphorus (P) and nitrogen (N) doping on the flame retardant performance of rGO/epoxy nanocomposites. The incorporation of PN-rGO increased char yield and reduced the mass loss rate of the composites due to the formation of a protective char layer. It was found that p- and n-doping increased the thermal-oxidative stability of rGO and promoted the dispersion of rGO in the epoxy matrix. Meanwhile, the doped groups catalysed the carbonization of epoxy chains and improved the char yield. The combination of catalysed carbonization and the high thermal stability of the rGO network promoted the formation of a strong char layer, which can reduce the oxidation of residual chars and restrict the formation of holes and cracks. Therefore, the peak heat release rate (PHRR) and total heat release (THR) of the composites were reduced by 31% and 29% with the addition of 5 wt% PN-rGO. It should be noted here that the heat release rate (HRR) is calculated by measuring the gas flow and oxygen concentration during combustion, while the THR is calculated by the integration of HRR versus time. In addition, the limiting oxygen index (LOI, the least oxygen concentration for the combustion of the polymer) was improved from 25 to 30%. The effects of the high thermal conductivity of graphene towards the flame retardancy properties are quite complex; a high thermal conductivity can increase the conduction of heat from the surface to the interior and thereby delay ignition while at the same time increase the peak heat release rate. Therefore, a balance between the effects of thermal conductivity and the shielding performance of the nanofillers should be pursued by adjusting the filler content, in order to minimise the heat release rate and maximise the time to ignition. An example of decreased PHRR of epoxy nanocomposites reinforced with GNPs has been reported as a result of the competition between the high thermal conductivity and the barrier effect of GNPs [169].

Apart from graphene, other 2D materials such as hBN have also been used as flame retardant additives in epoxy nanocomposites [170-173]. It has been shown that the oxidation temperature of BN nanosheets (700-850 °C) is much higher compared to that of graphene (450-500 °C) [30], which means that BN nanosheets can outperform graphene in the enhancement of thermal stability. Yu *et al.* [170] studied the effects of hydroxylated h-BN (BNO) on the thermal stability and flame retardancy of epoxy nanocomposites. The glass transition temperature of the epoxy was increased from 130.3 to 173 °C with the addition of 3 wt% BNO, as the covalently functionalized BNO nanosheets increased the crosslinking density and confined the movement of the molecular chains of the epoxy resin. In addition, the PHRR and THR of epoxy nanocomposites were decreased by 53% and 33% with the incorporation of 3 wt% BNO, which was attributed to the formation of a thermally stable char layer from BNO nanosheets. The char layer delayed the permeation of oxygen and reduced the release of combustible volatiles, resulting to the enhancement of flame retardancy.

Transition metal dichalcogenides such as $MoS_2$ are also halogen-free flame retardants and can reduce the fire hazards from epoxy resins [174-179]. For instance, a reduction of 43% in PHRR and 14.6% in THR of epoxy nanocomposites have been observed with the incorporation of 2 wt% chitosan-modified $MoS_2$ [174]. The enhancement was attributed to the physical barrier effect of $MoS_2$. The char layer can delay the release of combustible gases and decrease the effusion of combustible volatiles. Moreover, $MoS_2$ has been widely used in combination with other fillers to further improve the flame retardancy of epoxy nanocomposites. For example, the PHRR and THR of epoxy resin were reduced by 66% and 34% with the incorporation of 2 wt% of a hybrid layered double hydroxide/$MoS_2$ filler [177].

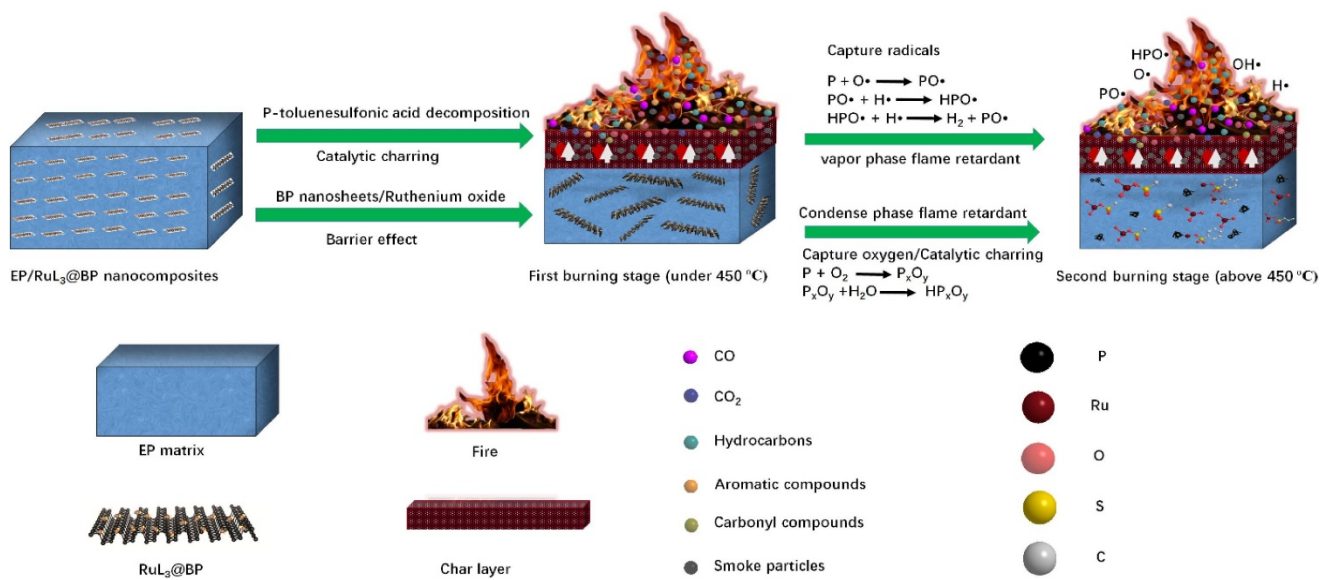

**Figure 11**. Flame retardancy mechanism of RuL$_3$@BP/epoxy nanocomposites. Reproduced with permission from [97]. Copyright 2020 Elsevier.

Another effective flame retardant additive is BP. Different from the additives which rely on the reinforcement in the condensed phase, the use of BP can affect both the gas phase and the condensed phase of flame retardancy simultaneously [97, 98, 180-182]. Therefore, BP can be considered a highly effective flame retardant in polymer nanocomposites. One drawback of BP lies in its instability under air, which has restricted its applications in polymer composites. Various modification methods have been proposed which can not only improve the air stability of BP nanosheets but also promote the dispersion of BP nanosheets in polymer matrices. Qu *et al*. [97] stabilized BP with $RuL_3$ via Ru-P coordination and studied the effects of functionalized BP on the flame retardancy of an epoxy resin. It was found that the incorporation of 3 wt% $RuL_3$@BP into the epoxy not only improved flame retardancy but also supressed the generation of smoke. The limiting oxygen index (LOI) was improved from 25% to 31%. In addition, the PHRR and THR were decreased by 62% and 35%, respectively. Factors that can be studied to assess the smoke generation, such as the smoke production rate (SPR), total smoke production (TSP) and the amount of CO produced per second (COP), were also significantly reduced. The flame retardancy mechanism is shown in **Figure 11**. During the combustion stage (< 450 ℃), the reactions between sulphur oxides and water molecules promote the formation of charring residues. The barrier effect of BP and ruthenium oxide delays the release of flammable gas and shields the underlying polymer from being burned. During the second stage (> 450 ℃), BP nanosheets reinforce the gas phase and the condensed phase simultaneously. The reactions between active radicals such as PO·, $PO_2$·, HPO· and highly reactive species such as H and OH generate less reactive products and inhibit the combustion process. Meanwhile, some BP nanosheets are oxidized into $P_xO_y$ and various phosphoric acid derivatives can react with epoxy to generate some new chemical bonds like O-P=O and P-O-P. These new chemical bonds accelerate the formation of char residuals, which can inhibit the generation of combustible gases, the transfer of heat and the emission of smoke. Therefore, both the gas phase and condensed phase are reinforced with the incorporation of BP nanosheets.

Other 2D materials such as MOFs and COFs [183-188] have also been used as flame retardant additives in epoxy nanocomposites. These materials show good compatibility with epoxy resins, without the need of any surface modifications. In addition, MOFs and COFs can be designed to contain nitrogen, phosphorus or aromatic rings to improve the flame retardancy of epoxy nanocomposites. Some of the most indicative 2D materials that have been used to improve the flame retardancy of epoxy resins and the peak heat release rate, total heat release along with the total smoke production and limiting oxygen index of the corresponding epoxy nanocomposites, are summarized in **Table 7**. Overall, BP is more effective in the reinforcement of flame retardancy considering its effectiveness towards both flame retardancy mechanisms (gas phase and condensed phase). Boron nitride and graphene are also excellent fire retardant additives for epoxies thanks to their high thermal stability.

**Table 7.** Flame retardancy of epoxy nanocomposites reinforced with 2D materials.

| Filler | Optimum filler fraction | PHRR $(kW/m^2)$/ Decrease (%) | THR $(mJ/m^2)$/ Decrease (%) | TSP $(m^2)$/ Decrease (%) | LOI (%)/ Increase (%) | Ref. |
|---|---|---|---|---|---|---|
| GNP | 3 wt% | 1796/47 | – | – | – | [163] |
| GNP | 3 wt% | –/-5 | 33.4/15 | – | 15.7/32 | [169] |
| f-rGO | 5 wt% | 1137/31 | 81.6/29 | 62.4/51 | 25/22 | [165] |
| f-GO | 3 wt% | 707/42 | 82.1/22 | – | – | [166] |
| rGO/hBN | 2/20 wt% | 1137/35 | 81.6/34 | 62.4/43 | 25/17 | [168] |
| hBN | 3 wt% | 1630/53 | 82.3/32 | 14.7/23 | – | [170] |
| f-BN | 12.1 vol% | 1197/42 | 82.7/38 | 66.8/53 | – | [171] |
| f-BN | 3 wt% | –/43 | –/48.3 | – | – | [173] |
| BN/ZF | 3 wt% | 1129/49 | – | – | 21.2/39 | [172] |
| f-MoS$_2$ | 2 wt% | 1592/43 | 39.7/14 | – | – | [174] |
| f-MoS$_2$ | 2 wt% | 1099/26 | 110/17 | 38/24 | – | [178] |
| MoS$_2$/MH | 2 wt% | 1183/32 | – | – | – | [175] |
| MoS$_2$/PZS | 3 wt% | –/41 | –/30 | – | – | [179] |
| MoS$_2$/LDH | 2 wt% | 1863/66 | 109/34 | 40/50 | – | [177] |

| f-BP | 1.2 wt% | –/43 | –/12 | – | 24.7/21 | [180] |
|------|---------|------|------|---|---------|-------|
| f-BP | 3 wt% | 1461/62 | 86.4/35 | 32.3/39 | 24.7/27 | [97] |
| BP/rGO | 1/1 wt% | 1462/55 | 105.6/54 | 54.1/28 | – | [181] |
| BP/MWCNT | 1/1 wt% | 2082/56 | 128.2/41 | 45/33 | – | [182] |
| BP/PZN | 2 wt% | 2116/59 | 167.1/63 | 50/30 | – | [98] |
| MOF | 2 wt% | 855/28 | 86/19 | – | – | [183] |
| MOF | 3 wt% | 1201/51 | 119/14 | 37.8/13 | 25.7/14 | [188] |
| GO/MOF | 9.5/0.5 wt% | 992/41 | – | 36.9/50 | 23.8/13 | [184] |
| COF | 3.2 wt% | 1369/18 | 135.6/18 | – | 23.5/6 | [185] |
| f-COF | 3.2 wt% | 1373/20 | – | – | 23.5/9 | [187] |
| GO/COF | 1.6/0.4 wt% | 1945/43 | 80/24 | – | – | [186] |

[a] ZF: Zinc ferrite; PZN: polyphosphazene.

*4.6 Anticorrosive properties*

The excellent resistance to chemicals, strong adhesion to substrates and high mechanical strength of epoxy resins render them suitable for anti-corrosion coatings ranging from heavy industry to daily life. An epoxy coating can be placed on the targeted metal to protect the metal from corrosion. For example, epoxy coatings have long been used to protect ships and marine structures. Metal cans and containers are often coated with epoxy resins to prevent rusting. Besides, epoxy resins have also been coated on paintings, sculptures, statues, souvenirs and laminate table tops. The use of epoxy coatings can provide a physical barrier effect and inhibit the penetration of corrosive media into metal surfaces. However, due to the effects of the corrosive environment (e.g. temperature, chemicals, solvents), pure epoxy coatings tend to deteriorate quickly after being exposed to corrosive media for a short time, given that they act as mixed-type corrosion inhibitors [189]. The anticorrosion mechanism of polymer coatings is shown in Figure 12a. The corrosion process can be divided into three steps: water absorption on the coating surface, diffusion of corrosive media in the coating matrix and final corrosion on the metal surface [190]. When fillers are well-dispersed in the matrix, the diffusion path of corrosive media is more tortuous and the barrier performance is improved. Various fillers have been incorporated into polymer coatings to improve their corrosion resistance and to prolong their service life [191]. Among these fillers, 2D nanosheets with their geometrical characteristics, large surface area and high gas/liquid barrier efficiency can outperform other fillers (e.g. with spherical and tubular shapes) in the improvement of anticorrosive performance as they can better distort the diffusion path of corrosive media and block corrosive media (e.g. water vapour) from entering the anticorrosion coating [192]. On that basis, a well-oriented network of fillers within a composite is able to outperform randomly dispersed systems. The incorporation of 2D nanosheets can also increase the hydrophobicity of the coating surface and increase the adhesion strength between the coating and the metal surface, which also lead to improved anticorrosion characteristics.

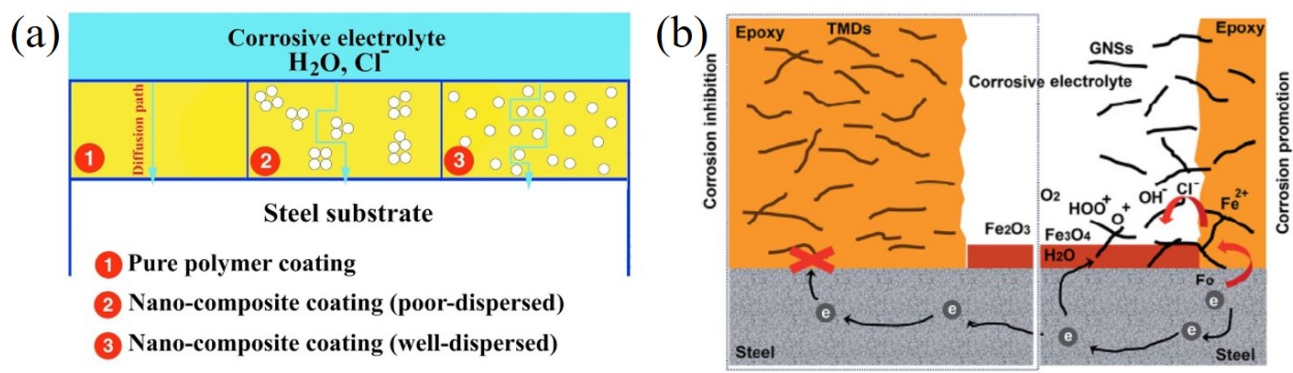

**Figure 12**. (a) Anticorrosion mechanism of polymer coatings with nanofillers. Reproduced with permission from [190]. Copyright 2020 Elsevier. (b) Corrosion mechanisms of epoxy nanocomposites with TMDs and graphene. Semiconducting TMDs can inhibit the electron transfer between the coating and the steel, while graphene with its high electrical conductivity allows the electron transfer. Reproduced with permission from [193]. Copyright 2019 The Royal Society of Chemistry.

Graphene has been widely used as an ultra-thin and lightweight anticorrosive additive in epoxy resins due to its large surface area and high impermeability [194-198]. Chen *et al*. [194] studied the effects of non-covalently functionalized graphene nanosheets on the corrosion behaviour of epoxy coatings. It was found that the P2BA-stabilized graphene improved the corrosion performance of epoxy nanocomposites as it reduced the water absorption of the epoxy after immersion in water for 10 days. The tortuous path created by well-dispersed GNPs prevents water from permeating through the epoxy. Additionally, the initial coating resistance was similar for the pure epoxy and epoxy nanocomposites (in the order of $10^9$ $\Omega$ cm$^2$). However, the coating resistance of pure epoxy decreased sharply after 80 days, while the corresponding value of epoxy nanocomposites was stable at $10^8$ $\Omega$ cm$^2$. The presence of functionalized graphene inhibited the penetration of electrolyte and improved the corrosive resistance of epoxy coatings. These results indicate that the use of graphene can improve the anticorrosive performance of epoxy coatings effectively. It should be pointed out that the high electrical conductivity of graphene restricts its application in long-term corrosion coatings (e.g. more than 80 days), as it can promote electrochemical reactions and galvanic corrosion [32]. The electrically insulating BN or semiconducting hexagonal MoS$_2$ can be effective alternatives to graphene for long-term anticorrosion applications.

Boron nitride nanosheets have also been investigated as barrier materials considering their chemical stability and high impermeability [199-202]. Additionally, boron nitride has an advantage towards enhancement of barrier performance compared to graphene, as its electrical resistivity can inhibit the galvanic corrosion under ambient environment. It has been shown that pure BN can outperform graphene in terms of long-term corrosion barrier due to its high impermeability and electrical insulating characteristics [32]. Therefore, it is expected that BN can enhance the anticorrosive performance of epoxy nanocomposites enduringly. Cui *et al*. [200] investigated the effect of BN on the anticorrosive performance of an epoxy coating. Poly(2-butylaniline) (PBA) was used once again to facilitate the exfoliation and dispersion of BN nanosheets and the corrosion behaviour of epoxy coatings was characterised. For the pure epoxy, the impedance modulus degraded dramatically, more than 3 orders of magnitude, after 50 days immersion. In contrast, the impedance modulus of the epoxy nanocomposites with 0.5 wt% BN remained stable after 120 days immersion. The addition of BN nanosheets inhibited the penetration of corrosive media by introducing a complex diffusion path for the corrosive media. In addition, the use of PBA passivates the metal surface and a protective layer is formed on the metal surface to block corrosive media, which ensures superior anticorrosive performance. In another study, it was found that functionalized $Al_2O_3$ can not only improve the stability and dispersion of BN but also promote the interfacial adhesion between BN nanosheets and epoxy matrix by acting as a spacer between the BN nanosheets [201]. Therefore, the diffusion path becomes even more complex and the anticorrosive performance of epoxy nanocomposites are significantly improved.

Compared to graphene, the low conductivity of TMDCs can reduce the electron transfer rate between coating and substrate and delay the galvanic corrosion of coatings significantly. Therefore, TMDCs are also expected to improve the anticorrosive performance of epoxy resins [193, 203, 204]. Ding *et al*. [193] compared the corrosion performance of epoxy nanocomposites reinforced by $MoS_2$, $WS_2$ and graphene as shown in **Figure 12**. The addition of $MoS_2$ and $WS_2$ reduced the oxygen and water vapour transmission rate due to the formation of a tortuous diffusion path for gas molecules. Additionally, it was found that $MoS_2$ and $WS_2$ provide better corrosion resistance in epoxy resins compared to graphene. For example, epoxy nanocomposites with 1 wt% $MoS_2$ and $WS_2$ displayed significantly better coating resistance than the epoxy samples reinforced with 1 wt% graphene. The reason for this, as mentioned earlier, is the high conductivity of graphene that allows electron transfer and formation of a conductive network between the coating and substrate, which can promote the galvanic corrosion of the coating. In contrast, the semiconducting $MoS_2$ or $WS_2$ can inhibit this process and improve the corrosion performance. Another disadvantage of graphene in such applications is its oxygen reduction catalytic activity, which can also promote the corrosion process. Overall, TMDCs can be considered more effective in the inhibition of corrosion compared to graphene.

MXenes [116, 118, 205] and MOFs [206-209] have also been used to enhance the anticorrosive performance of epoxy coatings. The results from a representative selection of works utilising epoxy/2D material anticorrosive coatings are outlined in **Table 8**. In summary, well-dispersed 2D materials with large diameter to thickness ratio and high impermeability are beneficial for the formation of tortuous diffusion pathways to inhibit the penetration of corrosive media in epoxy coatings. Additionally, 2D materials with low electrical conductivities such as BN and TMDCs display obvious advantages towards inhibiting the galvanic corrosion of epoxy coatings.

**Table 8.** Anticorrosive properties of epoxy nanocomposites reinforced with 2D materials (the corrosive medium in all works is a 3.5 wt% NaCl solution).

| Filler | Optimum filler fraction | Immersion time (days) | Impedance modulus (0.01 Hz, $\Omega$ cm$^2$) | Ref. |
|---|---|---|---|---|
| f-Gr | 0.5 wt% | 0, 5 | $8.25 \times 10^5$, $2.46 \times 10^5$ | [195] |
| f-GO | 0.5 wt% | 55 | $10^8$ | [196] |
| GO/PDA | – | 0, 40 | $1.6 \times 10^8$, $2.46 \times 10^7$ | [197] |
| Gr/P2BA | 0.5/0.5 wt% | 0, 80 | $7.45 \times 10^9$, $2.71 \times 10^8$ | [194] |
| GO/ZIF-8 | – | 2, 42 | $10^{11}$, $10^9$ | [198] |
| hBN | 1 wt% | 0, 10 | $9.8 \times 10^6$, $1.4 \times 10^6$ | [199] |

| hBN/PBA | 0.5 wt% | 2, 120 | $8.21\times10^9$, $1.51\times10^9$ | [200] |
|---|---|---|---|---|
| hBN/PEI | 0.3 wt% | 1, 70 | $2.45\times10^9$, $6.67\times10^7$ | [202] |
| BN/PDA/f-Al$_2$O$_3$ | 1 wt% | 1, 10 | $1.839\times10^{10}$, $3.387\times10^9$ | [201] |
| MoS$_2$/PDA | 0.5 wt% | 20 | $10^{10}$ | [203] |
| MoS$_2$ | 1 wt% | 60 | $8.1\times10^{10}$ | [193] |
| WS$_2$ | 1 wt% | 60 | $9.1\times10^{10}$ | [193] |
| Ti$_3$C$_2$ | 1 wt% | 0, 4 | $6.23\times10^8$, $2.96\times10^7$ | [205] |
| f-Ti$_3$C$_2$T$_x$ | 0.5 wt% | 0, 28 | $3.09\times10^9$, $1.02\times10^7$ | [116] |
| Ti$_3$C$_2$/Gr | 0.25/0.25 wt% | 0 | $2.14\times10^9$ | [118] |
| f-MOF | 0.5 wt% | 0 | $3.18\times10^8$ | [206] |
| f- ZIF-8/GO | 0.5 wt% | 0, 20 | $1.17\times10^{11}$, $3.66\times10^9$ | [208] |
| ZIF-8/SiO$_2$ | 2 wt% | 1, 30 | $10^{10}$, $1.21\times10^9$ | [207] |
| ZIF-8/ZnG | – | 0, 14, 50 | $10^7$, $10^{8.3}$, $10^{7.2}$ | [209] |

[a] PDA: polydopamine; PEI: poly-ethyleneimine; ZnG: zinc gluconate; ZIF-8: zeolitic imidazolate framework-8

## 4.7 Other properties

Apart from the above mentioned properties and applications, epoxy nanocomposites reinforced with 2D materials have been applied in other areas such as shape memory composites [210-218], biodegradable or biocompatible materials [219, 220] and biomedical systems [221, 222].

**Shape memory properties:** SMPs are able to retain a shape and switch to their permanent shape subjected to external stimulus such as heat, electricity, magnetic field, pH and light radiation [223]. However, the low thermal conductivity, the insulating nature and the low mechanical strength of epoxy resins have largely restricted their application as shape memory materials. Various 2D materials can be incorporated into an epoxy matrix to increase its thermal or electrical response, and therefore shape memory performance. As shown by the majority of literature reports, the shape memory performance is commonly obtained at low filler contents (< 2 wt%), indicating the scalability and the cost-effectiveness of the nanocomposite materials. For example, the mechanical properties and shape memory effects of graphene nanoplatelets in an epoxy were studied by Zhao *et al.* [213, 214] and results showed that the presence of GNPs (at 1 wt%) induced enhanced recovery stress and speed compared to neat epoxy. Similarly, Yu *et al.* [215] revealed that only 0.8 wt% of GO can lead to good shape recovery ratio, shape restoring rate and thermomechanical performance. In the interesting work of D'Elia *et al.* [216] Joule heating was utilised to induce a shape memory effect (shape fixity above 0.95 and shape recovery ratio above 0.98) for rGO/epoxy nanocomposites at low voltages and low filler contents (<1 wt%). The multifunctionality of the material allows damage sensing, while the filler network is able to divert crack propagation and increase fracture resistance. In another work, a hybrid GO/CNT filler was utilised to enhance filler dispersion and composite mechanical properties and conductivity, which led to a significant increase of the thermal response speed of epoxy nanocomposites [211]. The thermal response speed was significantly affected by the thermal conductivity at lower filler contents (< 2 wt%), while for higher filler loadings the storage modulus was the dominating factor, restricting the response of shape memory effect. The aqueous suspension route for the eco-friendly fabrication of the nanocomposites, along with the excellent shape memory performance makes the specific materials interesting for use in many areas such as textiles, sensors, aerospace/automotive industries and others. Finally, nacre-like shape memory rGO/epoxy composites have been shown to also display interesting shape memory properties [224]. The nacre-like composites displayed a shape memory effect with thermal and electrical stimuli, while the fracture toughness of the matrix was improved significantly (~2.5 times) at only 0.6 wt% of rGO content. The mechanism of shape memory for such composites (Figure 13a), involves the increase of temperature above the $T_g$, where the composites become soft (Figure 13a$_{1-2}$) and the subsequent application of stress, which changes chain conformations and results in a new, fixed shape (Figure 13a$_3$). The reduction of temperature leads to chain freezing and storage of entropic energy (Figure 13a$_4$); however, with the increase of temperature as a result of heat or electrical current (Figure 13a$_5$), the molecular chains are driven back to their lowest energy configuration (initial shape). Even though the literature on SMP/2D material nanocomposites is dominated by graphene-related materials, also MXenes could be used in a similar manner for the creation of electrically and thermally conductive SMP composites, while boron nitride and its hybrids [225-229] could also offer an alternative. Overall, it can be concluded that the introduction of 2D fillers into shape memory epoxy resins at low filler contents can effectively lead to strong self-shaping structures and can avoid catastrophic failure. The architecture of the nanocomposites needs to be fully controlled to take advantage of the 2D geometry of the fillers and their exceptional intrinsic properties for the creation of structural and multifunctional shape memory nanocomposites for various high-end applications.

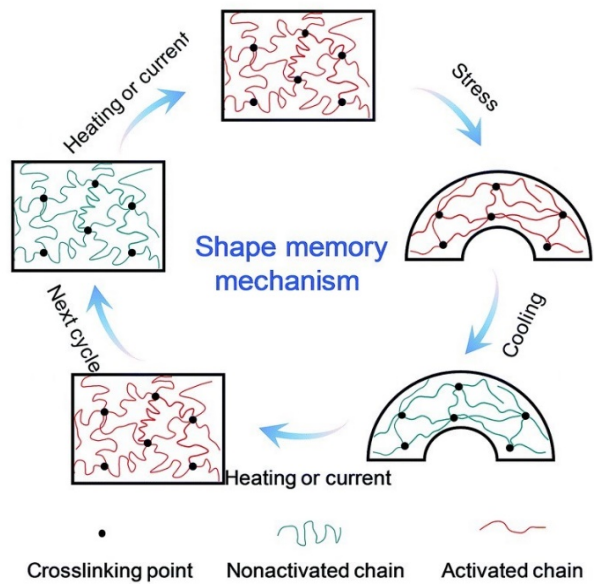

**Figure 13.** Shape memory mechanism for epoxy/2D material nanocomposites: (1) permanent shape of nanocomposite, (2) heating or current leads to activation of frozen chain segments, (3) application of stress leads to deformation of the nanocomposite, (4) cooling leads to a stress-free temporary shape

and (5) upon heating or re-exposure to current, the material recovers its permanent shape. Reproduced from ref. [224] with permission from the Royal Society of Chemistry..

**Biodegradability:** Traditional epoxy resins are petroleum-based and non-biodegradable and as a result significant attention has been paid from both academia and industry towards the fabrication of bio-based and biodegradable epoxy resins [230, 231]. The introduction of 2D fillers into biodegradable resins is a method to impart multifunctionality; however, the effect of the presence of nanomaterials within this type of resins has not been extensively studied. In the report of Baruah and Karak [220], the authors showed that the presence of GO in a bio-based hyperbranched epoxy resin at low contents (<0.5 wt%) enhanced biodegradability against both gram positive and negative bacteria. There is plenty of work to be done in the upcoming years to understand the effect of 2D materials on bio-based and biodegradable epoxy resins, from curing kinetics and reinforcing efficiency to the biodegrading mechanisms of epoxy resins. Clearly, the development of an eco-friendly future is certainly a good motivation for researchers across the globe.

## 5. Conclusions and outlook

Epoxy nanocomposites have been identified from several sectors/industries as excellent lightweight, structural and multifunctional materials for numerous applications. Two-dimensional materials, with their unique geometry and exceptional intrinsic properties can serve as extremely effective and promising multifunctional reinforcements of epoxies. A large number of literature works have reported significant improvements on the mechanical, thermal, electrical and EMI properties, along with excellent tribological, fire retardancy and anticorrosive capabilities.

In the present review, we have shown that the possibilities of 2D nanoreinforcements for advanced engineering applications are numerous as a result of the progress in nanocomposite manufacturing techniques as well as their multifunctionality. The most effective processing routes of epoxy nanocomposites with 2D materials have been identified, while the importance of a homogeneous dispersion, strong matrix-filler interactions and a possible controlled orientation of the fillers towards property enhancement has been highlighted throughout the manuscript.

The current literature status has been analysed thoroughly to identify and evaluate the critical parameters and inherent properties that affect the reinforcing efficiency of 2D materials within epoxy resins, while different 2D nanofillers ranging from graphene, boron nitride, molybdenum disulphide, MXenes, black phosphorus and others have been examined. Overall, graphene is quite expectedly the most popular nanofiller that has been used to reinforce epoxies, since it can produce nanocomposites with desired properties. However, epoxy-graphene nanocomposites are not free from disadvantages: the increase of graphene thickness or reduction of lateral size can reduce its mechanical effectiveness, while its high electrical conductivity can prevent its use in nanocomposites that demand electrical insulation or epoxy coatings where galvanic corrosion is undesired. For such composites, boron nitride is an alternative nanofiller, given that its mechanical properties do not degrade up to ten layers thick, while it is also an electrical insulator. Molybdenum disulphide can provide excellent tribological properties as a result of its weak interlayer interactions, while black phosphorus is the most effective additive for fire retardancy given that it can reinforce both flame retardancy mechanisms (gas phase and condensed phase).

A number of challenges and opportunities still exist for the mass production of epoxy nanocomposites with 2D materials, even though several attempts are under industrial and commercial investigation. Initially, the scale up of the production of affordable and high-quality 2D nanofillers should be the main target of manufacturers, as this is always going to be reflected on the ultimate properties of the nanocomposites. Nanoplatelets with large lateral dimensions, low thickness and low defect density should be the ideal platform to maximise reinforcement. The scale up of the fabrication of 2D and 3D structures of 2D materials and the subsequent successful impregnation of an epoxy can also offer many possibilities for achieving high-performing epoxy nanocomposites. The significant increase in the viscosity of epoxies can pose a problem for 2D nanomaterials that include functional groups such as graphene oxide, where the strong bonding between the matrix and the filler leads to a saturation of the mechanical properties at relatively low filler contents; additional research towards this direction is needed. For the case of BP nanocomposites, its stability under air is still a major problem and new routes can be explored for the facile preparation of BP-epoxy nanocomposites. Novel functionalisation routes should be also explored for the maximisation of reinforcing efficiency of 2D materials. Given that the formation of a strong polymer/filler interface is the key to realising the properties of 2D materials into composites, further research on the topic can provide the community with a better understanding on how interfacial interactions are affecting each property. Additionally, research onto the use of 2D materials in hierarchical fibre-reinforced composites should also offer a new venue of smart applications, while the use of hybrid fillers that combine continuous fibres with 2D nanofillers should be attractive for industries such as the aerospace that are already using fibre-reinforced plastics extensively. Finally, in line with the recent global initiatives towards the creation of bio-based and biodegradable polymers, the effect of 2D fillers on the biodegradability and the carbon footprint of novel epoxy/2D nanocomposites needs to be thoroughly understood. Additionally, aspects such as recyclability of novel epoxy resin nanocomposites and the biocompatibility of such components are expected to attract significant attention for the following years. Nevertheless, epoxy systems reinforced with 2D nanomaterials are envisaged to constitute the next-generation of engineering materials that will take advantage of the unique multifunctionality of 2D fillers and will offer important solutions for both academia and high-end industries.

**Acknowledgements**:

The authors acknowledge support from "Graphene Core 3" GA: 881603 which is implemented under the EU-Horizon 2020 Research & Innovation Actions (RIA) and is financially supported by EC-financed parts of the Graphene Flagship. MD is grateful to the China Scholarship Council for financial support.

Characterization of 2D Molybdenum Carbide (MXene). Adv Funct Mater. 2016;26:3118-27.

7.

MXene (Ti(3)C(2)Tx) for non-hydrophilic epoxy resin-based composites with enhanced mechanical and physical properties. Mater Des. 2021;197:11.